\documentclass[10pt,twocolumn,twoside]{IEEEtran}

\usepackage[T1]{fontenc}
\usepackage[utf8]{inputenc}
\usepackage{multirow}
\usepackage{tikz,ifthen,calc,graphicx,pgfplots,grffile}
\usepackage{amsmath}
\usepackage{amsfonts,amssymb}
\usepackage{hyperref}
\usepackage{cite}
\usepackage{mathrsfs}
\usepackage{color,bm}
\usepackage{algorithm}
\usepackage{algpseudocode}
\usepackage[mode=buildnew]{standalone}
\usepackage{stfloats}
\usepackage{subfig}

\usetikzlibrary{positioning,plotmarks,arrows.meta,spy}
\pgfplotsset{compat=newest}
\usepgfplotslibrary{patchplots}

\newtheorem{lemma}{Lemma}

\def\QED{{\setlength{\fboxsep}{0pt}\setlength{\fboxrule}{0.2pt}\fbox{\rule[0pt]{0pt}{1.3ex}\rule[0pt]{1.3ex}{0pt}}}}

\def\A{{\mathbf A}}
\def\B{{\mathbf B}}
\def\C{{\mathbf C}}

\def\X{{\mathbf X}}
\def\Y{{\mathbf Y}}
\def\R{{\mathbf R}}

\def\Z{{\mathbf Z}}
\def\I{{\mathbf I}}

\def\F{{\mathbf F}}
\def\E{{\mathbf E}}
\def\U{{\mathbf U}}
\def\V{{\mathbf V}}
\def\G{{\mathbf G}}

\def\S{{\mathbf S}}

\def\M{{\mathbf M}}

\def\w{{\mathbf w}}
\def\v{{\mathbf v}}
\def\x{{\mathbf x}}

\def\y{{\mathbf y}}

\def\u{{\mathbf u}}
\def\h{{\mathbf h}}

\def\b{{\mathbf b}}

\def\n{{\mathbf n}}

\def\0{{\mathbf 0}}
\def\1{{\mathbf 1}}

\def\Sigmab{{\boldsymbol \Sigma}}

\def\Psib{{\boldsymbol \Psi}}

\def\Thetab{{\boldsymbol \Theta}}

\def\Gammab{{\boldsymbol \Gamma}}

\def\Xib{{\boldsymbol \Xi}}

\DeclareMathOperator{\tr}{tr}

\DeclareMathOperator{\diag}{diag}
\DeclareMathOperator{\blkdiag}{blkdiag}

\title{Passive detection of a random signal common to multi-sensor reference and surveillance arrays}
\author{David~Ram\'irez,~\IEEEmembership{Senior~Member,~IEEE,}
        Ignacio~Santamaria,~\IEEEmembership{Senior~Member,~IEEE,}
        and Louis~L.~Scharf,~\IEEEmembership{Life~Fellow,~IEEE}
\thanks{D.~Ram\'irez is with the Department of Signal Theory and Communications, Universidad Carlos III de Madrid, Legan{\'e}s, Spain and with the Gregorio Mara\~n\'on Health Research Institute, Madrid, Spain (e-mail: david.ramirez@uc3m.es).}
\thanks{I.~Santamaria is with the Department of Communications Engineering, Universidad de Cantabria, Santander, Spain (e-mail: i.santamaria@unican.es).}
\thanks{L.~L.~Scharf is with the Department of Mathematics, Colorado State University, USA (e-mail: louis.scharf@colostate.edu).}
\thanks{The work of D. Ram{\'\i}rez was partially supported by MCIN/AEI/10.13039/501100011033/FEDER, UE, under grant PID2021-123182OB-I00 (EPiCENTER) and by the Office of Naval Research (ONR) Global under contract  N62909-23-1-2002. The work of I. Santamaria was funded by MCIN/AEI/10.13039/501100011033, under Grant PID2022-137099NB-C43 (MADDIE). The work of L. L. Scharf was supported by the Office of Naval Research (ONR) under contract N00014-21-1-2145 and the Air Force Office of Scientific Research (AFOSR) under contract FA9550-21-1-0169. This paper was presented in part at the 2023 IEEE International Conference on Acoustics, Speech, and Signal Processing (ICASSP 2023).}}

\begin{document}

\maketitle

\begin{abstract}
This paper addresses the passive detection of a common rank-one subspace signal received in two multi-sensor arrays. We consider the case of a one-antenna transmitter sending a common Gaussian signal, independent Gaussian noises with arbitrary spatial covariance, and known channel subspaces. The detector derived in this paper is a generalized likelihood ratio (GLR) test. For all but one of the unknown parameters, it is possible to find closed-form maximum likelihood (ML) estimator functions. We can further compress the likelihood to only an unknown vector whose ML estimate requires maximizing a product of ratios in quadratic forms, which is carried out using a trust-region algorithm. We propose two approximations of the GLR that do not require any numerical optimization: one based on a sample-based estimator of the unknown parameter whose ML estimate cannot be obtained in closed-form, and one derived under low-SNR conditions. Notably, all the detectors are scale-invariant, and the approximations are functions of beamformed data. However, they are not GLRTs for data that has been pre-processed with a beamformer, a point that is elaborated in the paper. These detectors outperform previously published correlation detectors on simulated data, in many cases quite significantly. Moreover,  performance results quantify the performance gains over detectors that assume only the dimension of the subspace to be known.
\end{abstract}


\section{Introduction}

Passive radar systems have garnered increasing attention over the past few decades \cite{Griffiths_passive_radar_1,Griffiths_passive_radar_2,Blum_receive_passive,passive_preamble,Hack_passive_noreference_signal,Nuria_passive_radar,DOA_passive_radar,passive_radar_automative}. They are a type of bistatic radar \cite{Willis2007} that operates without control over the transmitted signals. These systems use non-cooperative transmitters, known as illuminators of opportunity, which can include terrestrial TV \cite{DVB_passive} and FM broadcast transmitters \cite{FM_passive}, mobile phone base transceiver stations (e.g., 4G and 5G systems), and communication or navigation satellites \cite{Willis2007}. The use of illuminators of opportunity is the key advantage of passive radar systems. Firstly, they enable covert operations, as the system does not disclose its location. Secondly, only the receiver needs to be deployed, which makes the system simple and  cost-effective. Thirdly, because the transmitter is not part of the system, passive radar is energy-efficient.

Using non-cooperative transmitters results in waveforms that are not optimized for detection, potentially resulting in passive radar systems with insufficient detection probability. To address this issue and ensure operational passive radars, it is common practice to incorporate a second channel, referred to as the reference channel. The reference and surveillance channels may be realized using beamforming or directional antennas \cite{Hack_passive_radar_MIMO_networks}. The surveillance channel measures the transmitted signal when there is a target to reflect it, while the reference channel always measures a linearly distorted version of the transmitted signal. There is a slight difference between passive radar and passive source localization. In passive radar, the reference array receives a transmitted signal of opportunity regardless of whether there is a target in the surveillance field. In passive source localization \cite{Scharf_PSL_first}, neither the reference nor surveillance array receives a  signal if there is no transmitted signal of opportunity, typically a transmitting radar. The work in \cite{Zaimbashi_unified_2021} presents a very interesting unifying framework that develops detectors under several assumptions regarding the transmitted signal and/or channel.

Over the last few years, researchers have extensively studied passive radar with a reference channel and proposed numerous detection algorithms \cite{Coherence_book,Colone_passive_2009,Himed_cross_correlation,Cui2015,Bialkowski2011,gogineni_passive_2018,Nadakuditi_passive,Nacho_passive_rank_1,Nacho_passive_rank_p,Wang_Asilomar,euspico17_passive}. In single-input single-output (SISO) channels, the standard detector is based on the cross-correlation between the signal received by the surveillance and reference channels. However, this detector is not optimal since the signal of the reference channel is contaminated by noise \cite{Himed_cross_correlation}. To address this issue, \cite{Cui2015} derived a generalized likelihood ratio test (GLRT) modeling the transmitted signals as either deterministic or stochastic (Gaussian) for single-antenna receivers. Multiple-input multiple-output (MIMO) channels have also been extensively investigated. For instance, \cite{Bialkowski2011} proposed an ad-hoc detector for passive detection based on the generalized coherence \cite{cochran_Trans_SP_95}. Another ad-hoc detector is presented in \cite{gogineni_passive_2018,Nadakuditi_passive}, which cross-correlates the right singular vectors of the data matrices at both channels. 

Researchers have also proposed detectors based on first principles. The GLRT for stochastic waveforms was derived in \cite{Nacho_passive_rank_1,Nacho_passive_rank_p,Wang_Asilomar,euspico17_passive} under several assumptions for the spatial correlation of the noise. In this case, the signal modulates the covariance matrix of the observations, which yields a second-order model \cite[Ch. 7]{Coherence_book}. In the case of first-order models, where the unknown signal appears in the mean of the observations \cite[Ch. 7]{Coherence_book}, the literature is much richer. One of the early contributions is the work in \cite{Hack_passive_radar_MIMO_networks}, which derived the GLRT for an unknown deterministic transmitted signal in spatially and temporally white noise, and known/unknown channels. GLRTs and Rao tests, all under a first-order model, have been derived in \cite{yu_novel_2023,javidan_target_2020,fazlollahpoor_rao_2020}. Moreover, researchers have also studied the case where some a priori information about the signal is available, i.e., it is sparse in the frequency domain \cite{zhang_passive_2019,zhang_sparsity-based_2019,zhang_multistatic_2017} or it has some known format \cite{liu_passive_2020}. Moreover, \cite{Himed_MAP_radar_2021} considers the case that a priori information is also available on the channel coefficients. This a priori information has also been considered in the case of second-order models. For instance, \cite{zhang_multistatic_2017-1} derived the GLRT for an autoregressive signal.

Passive radar systems with multiple illuminators and/or receivers have also been considered in the literature based on the GLRT or the Rao test \cite{chalise_glrt_2018,zaimbashi_multistatic_2020,liu_passive_2021,fazlollahpoor_passive_2021,zaimbashi_multistatic_2022}. All previous works focus on detection in passive radar, but there is also a large body of literature that considers localization and tracking, see, for instance \cite{pak_target_2019,wang_signal_2018,wang_joint_2019,chen_direct_2023}, or more recently integrated sensing and communications \cite{chalise_performance_2017}.

In this paper, we address the detection of a Gaussian signal transmitted by a single-antenna illuminator of opportunity. That is, we consider a second-order model for measurements. This work assumes independent noises at the surveillance and reference arrays, each with arbitrary spatial covariance. Previous works in passive sensing under second-order models assumed that the channels are unknown, showing that the GLRT in this scenario is a monotone transformation of the canonical correlations between the surveillance and reference channel observations \cite{Nacho_passive_rank_1,Nacho_passive_rank_p,Wang_Asilomar,euspico17_passive}. In this paper, we focus on the case where the channels are known except for a scaling factor, i.e., a basis for the channels is known. When the channel subspaces are known, which can be the case for known array geometries, e.g., uniform linear arrays with known direction of arrival, the detection problem is more challenging and it has not yet been studied in the literature. In this paper, detectors for the case where dimension-one subspace channels are known are compared with detectors for the case where only the dimension of the subspace channel is known. It is shown that knowledge of the subspace channel, and use of this knowledge in a detector, brings significant increases in detection probability. Additionally, we show that our detectors also provide detection improvements compared to other detectors, such as the cross-correlation based detector \cite{Himed_cross_correlation} and the detector in \cite{gogineni_passive_2018,Nadakuditi_passive}, which computes the cross-correlation between the main right singular vectors of the data matrices at both channels.

The proposed detector employs the generalized likelihood ratio test, where all but one of the unknown parameters under both hypotheses are replaced with their maximum likelihood (ML) estimates.  The ML estimate of the covariance matrix of the reference channel  does not admit a closed-form solution. As an extension of our preliminary work in \cite{Rank_one_known_ICASSP}, where we proposed a sample-based estimator, in this work we obtain the ML estimate in closed-form of part of this matrix. For the remaining part, which is its normalized first column, we resort to numerical optimization. In particular, we use a trust-region algorithm \cite{MATLAB_fminunc} to maximize the compressed log-likelihood, which yields the GLRT. As an alternative to this optimization, we also consider a sub-optimum estimator of the reference covariance using only a sample covariance matrix. The resulting detector we call the sample-GLR. We also propose a closed-form detector in the low-signal-to-noise (SNR) regime, which can be seen as a low-SNR GLR. It has an insightful interpretation as the correlation coefficient between the output of Capon beamformers. The performance of these three  detector statistics is evaluated through Monte Carlo simulations and compared with that of \cite{Nacho_passive_rank_1,Nacho_passive_rank_p}, which is the GLR for channel subspaces known only by their dimension, the cross-correlation based detector \cite{Himed_cross_correlation}, and the detector based on cross-correlating the singular vectors \cite{gogineni_passive_2018,Nadakuditi_passive}. From a practical stand-point, the sample-based estimator achieves a good trade-off between performance and computational complexity in a variety of scenarios.

\subsection{Outline}

The structure of this paper is as follows. Section \ref{sec:formulation} presents the second-order passive detection problem for Gaussian signals in known subspaces, which is formulated as a test for the covariance structure of the observations. Section \ref{sec:detector} derives the GLR, Section \ref{sec:simplfied} derives the two approximate GLRs, and Section \ref{sec:simulations} evaluates their performance through simulations. Section \ref{sec:conclusions} summarizes the main conclusions of this work.

\subsection{Notation}

Throughout this paper, matrices are denoted by bold-faced uppercase letters, column vectors by bold-faced lowercase letters, and scalars by light-face lowercase letters. The superscripts $(\cdot)^T$ and $(\cdot)^H$ represent the transpose and Hermitian transpose, respectively, and $\A \succ \0$ means that $\A$ is positive definite. The trace and determinant of a matrix $\A$ are denoted by $\tr(\A)$ and $\det(\A)$, respectively. The set $\mathbb{C}^{L}$ denotes the complex Euclidean space with the standard inner product. The notation $\text{blkdiag}_L(\A)$ denotes a block-diagonal matrix constructed from the $L \times L$ blocks in the diagonal of $\A$. Additionally, we use the notation $\x \sim \mathcal{CN}_{L}(\0, \R)$ to denote a proper complex Gaussian vector in ${\mathbb{C}}^L$ with zero mean and covariance $\R$. Finally, $\Re(\cdot)$ returns the real part of its argument.

%%%%%%%%%%%%%%%%%%%%%%%%%%%%%%%%%%%%%%%%%%%%%%%
\section{The passive detection problem}
\label{sec:formulation}

\begin{figure}[!t]
	\centering
	\includegraphics[width=\columnwidth]{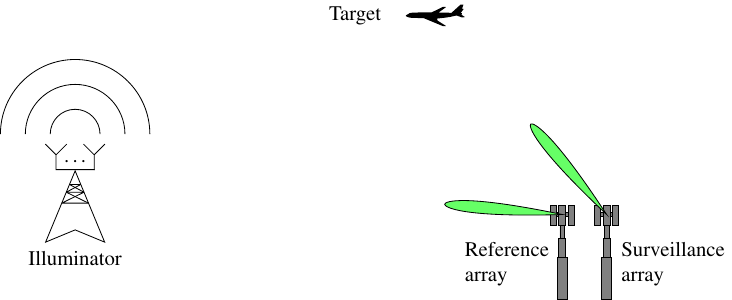}
	\caption{Schematic diagram of a passive radar}
	\label{fig:diagram}
\end{figure}

Passive detection addresses the problem of deciding whether a signal is present when there is no control over the transmitted signal. Typically, to improve performance, passive detection is  implemented using two multi-sensor arrays, as depicted in Fig. \ref{fig:diagram}. In one of the arrays, known as the \emph{reference} array, the transmitted signal is always present. In the so-called \emph{surveillance} array, the transmitted signal is observed only when a reflecting target is present. In this work, we consider the detection of a signal transmitted by a single-antenna illuminator of opportunity, in the presence of independent noises with arbitrary spatial covariances. Moreover, we assume that the surveillance signal is synchronized in delay and Doppler with the reference signal, which is achieved via a scanning process in the range-Doppler plane. This actually makes our proposed statistics ambiguity scores. This work also assumes that the direct-path signal has been removed by directional antennas or signal processing algorithms \cite{Colone_passive_2009}. When the power of direct-path leakage through sidelobes of a beamformer is significant, then this may be modeled as in \cite{Zhao_linear_fusion_2017,zhang_multistatic_2017,Zhang_delay_Doppler_direct_2016,Ramirez_MFA_2020}. For purposes of this paper, pre-processing in delay and Doppler produces a detector peak when the delay and Doppler match the delay and Doppler of the target signal and when they match the delay and Doppler of the direct path signal. This latter peak can be identified as a direct-path peak by a priori information about the coordinates of the illuminator with respect to the surveillance platform.

Given the previous assumptions, and considering $L$ receiving antennas in both arrays,\footnote{The generalization to a number of antennas different at each array is straightforward.} the signal model is
\begin{equation}
\label{eq:detection_problem}
\begin{array}{ll}
\mathcal{H}_0: \y_{n} = \begin{bmatrix} \0 \\ \h_r \end{bmatrix}  x_{n} +  \begin{bmatrix} \n_{s,n} \\ \n_{r,n} \end{bmatrix}, \vspace{0.2cm}
\\
\mathcal{H}_1: \y_{n} = \begin{bmatrix} \h_s \\ \h_r \end{bmatrix} x_{n} + \begin{bmatrix} \n_{s,n} \\ \n_{r,n} \end{bmatrix},
\end{array}
\end{equation}
where $\y_{n} = [ \y_{s,n}^T \, \y_{r,n}^T]^T \in \mathbb{C}^{2 L}$, $\y_{s,n} \in \mathbb{C}^{L}$ is the received signal at the surveillance channel $\h_s  \in \mathbb{C}^{L}$, and $\y_{r,n} \in \mathbb{C}^{L}$ is the received signal at the reference channel $\h_r  \in \mathbb{C}^{L}$. Moreover, $x_{n}$ is the transmitted signal and $\n_{n} = [ \n_{s,n}^T \, \n_{r,n}^T]^T \in \mathbb{C}^{2 L}$ stacks the independent noises at the surveillance array and at the reference array, which are distributed as $\n_{s,n} \sim \mathcal{CN}_L(\0, \Sigmab_{ss})$ and $\n_{r,n} \sim \mathcal{CN}_L(\0, \Sigmab_{rr})$, with the $L \times L$ covariance matrices $\Sigmab_{ss}, \Sigmab_{rr}$ constrained to be only positive definite, i.e., they are not structured as scaled identity or diagonal matrices.

In this work, we consider a scenario where the surveillance and reference channels subspaces $\langle \h_s \rangle, \langle \h_r\rangle$, are known. This is the case if we consider a uniform linear array (ULA) with known direction of arrival (DOA), or for unknown DOA when there is an angle grid and one point on the grid is considered. The same idea may be applied to a uniform circular array or to any known geometry that allows us to compute a parametric array manifold. With this assumption, we may decompose the channels as $\h_s = a_s \u_s$ and $\h_r = a_r \u_r$, where $\u_s, \u_r$ are the known unit-length basis vectors for the surveillance and reference channel subspaces, and $a_s, a_r$ are unknown complex amplitudes, which model unknown path loss, fading gain, and carrier phase.

To further proceed, we need some assumptions on the transmitted signal. In previous works, it is assumed that the signal is an unknown deterministic parameter, which yields a first-order model, where the information about $x_{n}$ appears in the mean of the observations \cite{first_vs_second_TSP,Coherence_book}. However, in this work, we solve for the GLR in the case where $x_{n}$ is distributed as $x_{n} \sim \mathcal{CN}(0, \sigma^2_x)$, with unknown variance $\sigma^2_x$. The joint distribution of the measurement and the transmitted signal is then integrated for the marginal distribution of the measurement. This results in a second-order  Gaussian model for the  measurement, where the covariance matrix for the signal 
appears in the covariance matrix of the observations.

Based on these assumptions, the detection problem is a test for the covariance structure of the observations:
\begin{equation}
\label{eq:detection_problem_covariance}
\begin{array}{ll}
\mathcal{H}_0: \y_{n} \sim \mathcal{CN}_L(\0, \R_0),
\\
\mathcal{H}_1: \y_{n} \sim \mathcal{CN}_L(\0, \R_1).
\end{array}
\end{equation}
It is a straightforward exercise to show that the covariance matrix $\R_0$ is
\begin{equation}
  \R_0 = \begin{bmatrix}
    \Sigmab_{ss} & \0 \\ \0 & \sigma_x^2 \h_r \h_r^H + \Sigmab_{rr}
  \end{bmatrix},
\end{equation}
and the covariance matrix $\R_1$ is
\begin{equation}
  \R_1 = \begin{bmatrix}
    \sigma_x^2 \h_s \h_s^H + \Sigmab_{ss} & \sigma_x^2 \h_s \h_r^H \\ \sigma_x^2 \h_r \h_s^H & \sigma_x^2 \h_r \h_r^H + \Sigmab_{rr}
  \end{bmatrix},
\end{equation}
where each of the four blocks of the covariance matrices is of dimensions $L \times L$. Alternatively, these covariance matrices can be written in terms of the channel unitary bases $\u_s, \u_r$ as \cite{Coherence_book}
\begin{equation}
  \R_0 = \begin{bmatrix}
     \Sigmab_{ss} & \0 \\ \0 & q_{rr} \u_r \u_r^H + \Sigmab_{rr}
  \end{bmatrix},
\end{equation}
and 
\begin{equation}
  \R_1 = \begin{bmatrix}
     q_{ss} \u_s \u_s^H + \Sigmab_{ss} & q_{sr} \u_s \u_r^H \\ q_{sr}^{\ast} \u_r \u_s^H & q_{rr} \u_r \u_r^H + \Sigmab_{rr}
  \end{bmatrix},
\end{equation}
where $q_{ss} = \sigma_x^2 |a_s|^2, q_{rr} = \sigma_x^2 |a_r|^2,$ and $q_{sr} = \sigma_x^2 a_s a_r^{\ast}$, are unknown parameters. Then, the unknown parameters $a_s,a_r,$ and $\sigma_x^2$ are transformed to $q_{ss}, q_{rr}$, and $q_{sr}$, with $q_{ss}, q_{rr}>0$, $q_{sr}=q_{ss}^{1/2}q_{rr}^{1/2}e^{j\phi}$, and $\phi \in [0, 2 \pi)$.

\section{Derivation of the generalized likelihood ratio}
\label{sec:detector}

In this section, the generalized likelihood ratio (GLR) for the detection problem in \eqref{eq:detection_problem_covariance} is derived, assuming that we have $N$ independent and identically distributed observations of $\y_{n}$, $\Y = [\y_1 \, \cdots \, \y_N] = [\Y_s^T \ \Y_r^T]^T,$ with $\Y_s = [\y_{s,1} \, \cdots \, \y_{s,N}]$ and $\Y_r = [\y_{r,1} \, \cdots \, \y_{r,N}]$. The GLR is defined as \cite{Kay_detection}
\begin{equation}
	\label{eq:GLR_definition}
	\Lambda = \frac{\displaystyle \max_{\R_1} \ell(\R_1 ; \Y )}{\displaystyle \max_{\R_0} \ell (\R_0 ; \Y)},
\end{equation}
where $\ell(\R_i ; \Y )$ is the likelihood of the $i$th hypothesis given by \cite{book_peter}
\begin{equation}
	\ell(\R_i; \Y ) = \frac{1}{\pi^{2 L N} \det(\R_i)^{N}} \exp \left\{-N\tr  \left( \R_i^{-1} \S \right)\right\},
\end{equation} 
with sample covariance matrix
\begin{equation}
	\S = \frac{1}{N} \Y \Y^H = \frac{1}{N} \sum_{n = 1}^{N} \y_n \y_n^H.
\end{equation}
The GLR is used in the GLR test (GLRT) \cite{Kay_detection}
\begin{equation}
	\label{eq:GLRT_definition}
	\Lambda \mathop{\gtrless}^{\mathcal{H}_1}_{\mathcal{H}_0} \mu.
\end{equation}

\subsection{Compressed likelihood under $\mathcal{H}_0$}

The compressed likelihood under $\mathcal{H}_0$ is %given by
\begin{equation}
	\max_{\R_0} \ell (\R_0 ; \Y) = \ell(\hat{\R}_0 ; \Y ),
\end{equation}
which requires the computation of the maximum likelihood (ML) estimate of $\R_0$. To compute this estimate, it is convenient to re-parametrize the covariance matrix as
\begin{equation}
	\R_0 = \begin{bmatrix}
		\R_{ss} & \0 \\ \0 & \R_{rr}
	\end{bmatrix},
\end{equation}
where $\R_{ss}, \R_{rr}$ are only constrained as $\R_{ss}, \R_{rr} \succ \0$ since $\Sigmab_{ss}, \Sigmab_{rr} \succ \0$. That is, the rank-one structure induced by the received signal in the reference channel disappears. Then, $\R_0$ is a block-diagonal positive definite matrix and its ML estimate is the block-diagonal version of the sample covariance obtained by zeroing out the Northeast (NE) and Southwest (SW) blocks \cite{mardia_multivariate_analysis}:
\begin{equation}
	\hat{\R}_0 = \blkdiag_L(\S) = \begin{bmatrix}
		\S_{ss} & \0 \\ \0 & \S_{rr}
	\end{bmatrix},
\end{equation} 
with the sample covariance matrix patterned as
\begin{equation}
	\S = \begin{bmatrix}
		\S_{ss} & \S_{sr} \\ \S_{sr}^H & \S_{rr}
	\end{bmatrix},
\end{equation}
where $\S_{ss}, \S_{sr}, \S_{rr} \in \mathbb{C}^{L \times L}$. Then, the compressed log-likelihood under $\mathcal{H}_0$ is
\begin{equation}
	\label{eq:loglikelihood_H0}
	% \log \ell (\hat{\R}_0 ; \Y) = - 2 L N (\log \pi + 1) - N \log \det(\S_{ss}) - N \log \det(\S_{rr}),
	\log \ell (\hat{\R}_0 ; \Y) = -  \log \det(\S_{ss}) - \log \det(\S_{rr}).
\end{equation}
In \eqref{eq:loglikelihood_H0}, we have omitted constant and multiplicative terms in the log-likelihood that do not depend on data, and this will be done in the remainder of the paper in all log-likelihoods.

\subsection{Compressed likelihood under $\mathcal{H}_1$}

In this subsection, we turn our attention to the ML estimation of the unknowns under $\mathcal{H}_1$, which will be much more challenging. In fact, we are able to derive in closed form the ML estimator functions for all unknowns but one $L$-dimensional unit-norm vector that can be easily optimized numerically. In particular, we will be able to obtain the closed-form ML estimates through several re-parametrizations of the problem. The first re-parametrization was originally used in \cite{Stoica_array_correlared}. First, and similarly to $\mathcal{H}_0$, since the noise covariance matrices are only constrained as $\Sigmab_{ss}, \Sigmab_{rr} \succ \0$, and with some abuse of notation, we rewrite $\R_1$ as
\begin{equation}
	\label{eq:R_1}
	\R_1 =  \begin{bmatrix}  \R_{ss} & q_{sr} \u_s \u_r^H \\  q_{sr}^* \u_r \u_s^H  & \R_{rr} \end{bmatrix},
\end{equation}
where $\R_{ss}$ and $\R_{rr}$ are again $L \times L$ positive definite matrices without further structure and $\R_{sr} = q_{sr} \u_s \u_r^H$ is an $L \times L$ rank-one matrix. We proceed with the block-UDL factorization \cite{Coherence_book} of \eqref{eq:R_1}:
\begin{multline}
	\R_1 =  \begin{bmatrix}  \I_L &  q_{sr} \u_s \u_r^H \R_{rr}^{-1} \\  \0  & \I_L \end{bmatrix} \times \\  \begin{bmatrix}  \R_{ss} - |q_{sr}|^2 \u_s (\u_r^H \R_{rr}^{-1} \u_r) \u_s^H &  \0 \\  \0  & \R_{rr} \end{bmatrix}  \begin{bmatrix}  \I_L &  \0 \\ q_{sr}^{\ast} \R_{rr}^{-1} \u_r \u_s^H  & \I_L \end{bmatrix},
\end{multline}
and define
\begin{align}
	\Thetab &= \R_{ss} - |q_{sr}|^2 \u_s (\u_r^H \R_{rr}^{-1} \u_r) \u_s^H,
\end{align}
which is the Schur complement of the block $\R_{rr}$ of $\R_1$. Since $\R_1 \succ \0$ and $\R_{rr} \succ \0$, the Schur complement is also positive definite, $\Thetab \succ \0$. With this definition, the covariance matrix becomes
\begin{multline}
	\R_1 =  \begin{bmatrix}  \I_L &  q_{sr} \u_s \u_r^H \R_{rr}^{-1} \\  \0  & \I_L \end{bmatrix} \times \\  \begin{bmatrix}  \Thetab &  \0 \\  \0  & \R_{rr} \end{bmatrix}  \begin{bmatrix}  \I_L &  \0 \\ q_{sr}^{\ast} \R_{rr}^{-1} \u_r \u_s^H  & \I_L \end{bmatrix}.
\end{multline}
Defining this Schur complement results in a re-parametrization that is a one-to-one mapping between the transformed parameters $\{\Thetab,\R_{rr},q_{sr}\}$ and the original parameters $\{\R_{ss},\R_{rr},q_{sr}\}$. The log-likelihood can be expressed in terms of the new parameters as
\begin{multline}
	\label{eq:loglikeli_Stoica}
	% \log \ell (\R_1 ; \Y) = \log \ell(\R_{rr}, \Thetab,q_{sr}; \Y ) =  - 2 L N \log \pi - N \log \det(\R_{rr}) \\ - N \tr \left(\R_{rr}^{-1} \S_{rr} \right) - N \log \det(\Thetab) - N \tr \left(\Thetab^{-1} \M(q_{sr},\R_{rr})), 
	\log \ell (\R_1 ; \Y) = \log \ell( \Thetab,\R_{rr},q_{sr}; \Y ) =  - \log \det(\R_{rr}) \\ 
-\log \det(\Thetab) - \tr \left(\R_{rr}^{-1} \S_{rr} \right)  -   \tr \left(\Thetab^{-1} \M(q_{sr},\R_{rr}) \right),
\end{multline}
where
\begin{multline}
	\label{eq:M_matrix}
	\M(q_{sr},\R_{rr}) = \S_{ss} + |q_{sr}|^2 \eta_r(\R_{rr}) \u_s  \u_s^H \\ -  q_{sr} \u_s \u_r^H \R_{rr}^{-1} \S_{sr}^H - q_{sr}^{\ast} \S_{sr} \R_{rr}^{-1} \u_r \u_s^H,
\end{multline}
and
\begin{equation}
	\eta_{rr}(\R_{rr}) = \u_r^H \R_{rr}^{-1} \S_{rr} \R_{rr}^{-1} \u_r.
\end{equation} 
As we will see, the new parametrization allows for a simpler maximization of the log-likelihood. The next lemma provides the ML estimates of $\Thetab$ and $q_{sr}$.
\begin{lemma}
	\label{lem:ML_Theta_b_q_sr}
	The ML estimates of $\Thetab$ and $q_{sr}$ are
	\begin{align}
		\hat{\Thetab} &= \M(\hat{q}_{sr},\R_{rr}),\\
		\hat{q}_{sr}(\R_{rr}) &= \frac{\eta_{sr}(\R_{rr})}{|\eta_{sr}(\R_{rr})|^2 + (\u_s^H \S_{ss}^{-1} \u_s) (\eta_{rs}(\R_{rr}) - \alpha_{sr}(\R_{rr}))}, \label{eq:q_sr_hat}
	\end{align}
where 
\begin{align}
	\eta_{sr}(\R_{rr}) &= \u_s^H \S_{ss}^{-1} \S_{sr} \R_{rr}^{-1} \u_r,\\
	\alpha_{sr}(\R_{rr}) &= \u_r^H \R_{rr}^{-1} \S_{sr}^H\S_{ss}^{-1}    \S_{sr} \R_{rr}^{-1} \u_r.
\end{align}
\end{lemma}

\begin{proof}
	See Appendix \ref{App:ML_Theta_b_q_sr}.
\end{proof}
\ \newline
The ML estimates in Lemma \ref{lem:ML_Theta_b_q_sr} yield the compressed log-likelihood
\begin{multline}
	\label{eq:log_likeli_Theta_b_qsr}
	% \log \ell(\R_{rr}, \hat{\Thetab},\hat{q}_{sr}; \Y ) =  -  L N (2 \log \pi + 1) - N \log \det(\S_{ss})  \\ - N \log \det(\R_{rr}) - N \tr \left(\R_{rr}^{-1} \S_{rr} \right) \\ + N \log \left(1 + \frac{|\eta_{sr}(\R_{rr})|^2}{\eta_s (\eta_r(\R_{rr}) - \alpha(\R_{rr}))}\right). \\
	% \log \ell(\R_{rr}, \hat{\Thetab},\hat{q}_{sr}; \Y ) =  -  L N (2 \log \pi + 1) - N \log \det(\S_{ss})  \\ - N \log \det(\R_{rr}) - N \tr \left(\R_{rr}^{-1} \S_{rr} \right) - N \log (\eta_s) \\ + N \log \left(\eta_s + \frac{|\eta_{sr}(\R_{rr})|^2}{\eta_r(\R_{rr}) - \alpha(\R_{rr})}\right).
	% \log \ell(\R_{rr}, \hat{\Thetab},\hat{q}_{sr}(\R_{rr}); \Y ) =  - \log \det(\S_{ss}) \\ - \log \det(\R_{rr}) - \tr \left(\R_{rr}^{-1} \S_{rr} \right) \\ + \log \left(1 + \frac{|\eta_{sr}(\R_{rr})|^2}{\eta_s (\eta_r(\R_{rr}) - \alpha(\R_{rr}))}\right). \\
	\log \ell(\hat{\Thetab},\R_{rr}, \hat{q}_{sr}; \Y ) =  - \log \det(\S_{ss}) - \log (\u_s^H \S_{ss}^{-1} \u_s)  \\ - \log \det(\R_{rr}) - \tr \left(\R_{rr}^{-1} \S_{rr} \right) \\ + \log \left(\u_s^H \S_{ss}^{-1} \u_s  + \frac{|\eta_{sr}(\R_{rr})|^2}{\eta_{rr}(\R_{rr}) - \alpha_{sr}(\R_{rr})}\right).
\end{multline}
The ML estimation of $\R_{rr}$, i.e., the maximization of \eqref{eq:log_likeli_Theta_b_qsr}, cannot be obtained in closed form, but it is possible to further compress the log-likelihood before resorting to numerical methods. To do so, we introduce a second re-parametrization. Define the unitary matrix $\U_r=[\u_r \ \V_r]$, where $\u_r^H\u_r=1$, $\u_r^H\V_r=\0$, and $\V_r^H\V_r=\I_{L-1}$.  Write $\R_{rr}^{-1}$ as
\begin{equation}
	\R_{rr}^{-1} = \U_r\X\U_r^H,
\end{equation}
where $\X \succ \0$ since $\R_{rr} \succ \0$. Moreover, $\X$ can be written as
\begin{equation}
	\X = \begin{bmatrix} \gamma w & \gamma \w^H \\ \gamma \w & \Z \end{bmatrix},
\end{equation}
where $\gamma > 0$ is the norm of the first column of $\X$ and $\x = [w \ \w^T]^T \in \mathbb{C}^{L}$ is the normalized (unit-norm) first column of $\X$. These new parameters are constrained as $\gamma, w > 0$, $\Z \succ \0,$ and $w - \gamma \w^H \Z^{-1} \w > 0$. Then, Appendix \ref{App:re_parametrization} derives the compressed log-likelihood in terms of the new parameters, which is given by
\begin{multline}
	\label{eq:log_likeli_Theta_b_qsr_reparametrized}
	% \log \ell(\R_{rr}, \hat{\Thetab},\hat{q}_{sr}; \Y ) = \log \ell(\gamma,\x,\Z, \hat{\Thetab},\hat{q}_{sr}; \Y ) =  -  L N (2 \log \pi + 1) - N \log \det(\S_{ss}) \\ - N \log (\eta_s) + N\log \gamma + N \log w + N\log \det \left(\Z - \frac{\gamma}{w} \w \w^H\right) \\  - N \gamma w \u_r^H\S_{rr} \u_r - 2 N \gamma \Re \left\{ \w^H \V_r^H\S_{rr} \u_r \right\} - N \tr\left(\Z \V_r^H \S_{rr} \V_r\right) \right)  \\ + N \log \left(\frac{ \x^H \Psib \x}{\x^H \Gammab \x}\right),
	\log \ell(\hat{\Thetab},\R_{rr}, \hat{q}_{sr}; \Y ) = \log \ell(\hat{\Thetab},\gamma,\x,\Z, \hat{q}_{sr}; \Y ) =   \\ 
- \log \det(\S_{ss}) - \log (\u_s^H \S_{ss}^{-1} \u_s)  + \log \gamma + \log w\\ 
+ \log \det \left(\Z - \frac{\gamma}{w} \w \w^H\right)   - \gamma w \u_r^H\S_{rr} \u_r - 2 \gamma \Re \left\{ \w^H \V_r^H\S_{rr} \u_r \right\} \\
 - \tr\left(\Z \V_r^H \S_{rr} \V_r\right)    + \log \left(\frac{ \x^H \Psib \x}{\x^H \Gammab \x}\right),
\end{multline}
where $\Psib$ and $\Gammab$ are defined in \eqref{eq:Psi_definition} and \eqref{eq:Gamma_definition}, respectively. The next lemma derives the ML estimates of $\Z$ and $\gamma$.

\begin{lemma}
	\label{lem:estimates_Z_gamma}
	The ML estimates of $\Z$ and $\gamma$ are
	\begin{align}
		\label{eq:ML_Z}
		\hat{\Z} &= (\V_r^H \S_{rr} \V_r)^{-1} + \frac{\hat{\gamma}}{w} \w \w^H,\\
		\label{eq:ML_gamma}
		\hat{\gamma} &= \frac{w}{\x^H \Xib \x},
	\end{align}
	where
	\begin{equation}
		\label{eq:Xi_definition}
		\Xib =\U_r^H \S_{rr} \U_r =  \begin{bmatrix}   \u_r^H\S_{rr} \u_r & \u_r^H\S_{rr} \V_r  \\  \V_r^H\S_{rr} \u_r & \V_r^H \S_{rr} \V_r  \end{bmatrix}.
	\end{equation}
\end{lemma}

\begin{proof}
	See Appendix \ref{App:estimates_Z_gamma}.
\end{proof}

The ML estimates in \eqref{eq:ML_Z} and \eqref{eq:ML_gamma} yield the compressed log-likelihood
	\begin{multline}
	\label{eq:last_compressed_likelihood}
	% \log \ell(\hat{\gamma},\x,\hat{\Z}, \hat{\Thetab},\hat{q}_{sr}; \Y ) =  - 2 L N ( \log \pi + 1) - N \log \det(\S_{ss}) \\ - N \log (\eta_s) - N\log \det \left(\V_r^H \S_{rr} \V_r\right) + N\log \left( \frac{\x^H \E \x}{\x^H \Xib \x}\right) \\ + N \log \left(\frac{ \x^H \Psib \x}{\x^H \Gammab \x}\right),
	\log \ell(\hat{\Thetab},\hat{\gamma},\x,\hat{\Z}, \hat{q}_{sr}; \Y ) =   - \log \det(\S_{ss}) - \log (\u_s^H \S_{ss}^{-1} \u_s) \\ - \log \det \left(\V_r^H \S_{rr} \V_r\right) + \log \left( \frac{\x^H \E \x}{\x^H \Xib \x}\right) + \log \left(\frac{ \x^H \Psib \x}{\x^H \Gammab \x}\right),
\end{multline}
where $\E = \diag(1, 0, \ldots, 0)$. The last step to obtain the compressed log-likelihood under $\mathcal{H}_1$ is the maximization of \eqref{eq:last_compressed_likelihood} with respect to $\x$. That is, maximize
	\begin{equation}
	\label{eq:cost_function}
	J(\x) =  \log \left( \frac{\x^H \E \x}{\x^H \Xib \x}\right) + \log \left(\frac{ \x^H \Psib \x}{\x^H \Gammab \x}\right).
\end{equation}
This cost function is not convex so we maximize it by resorting to iterative methods. Since the cost function can be expressed as a linear combination of logarithms of quadratic forms, it is straightforward to compute its gradient and Hessian. Moreover, the cost function is invariant to $\|\x\|$ and to the phase of the first element of $\x$, which is constrained as $w > 0$. 

Finally, this numerical procedure requires an initialization. Since
\begin{equation}
	 \X = \begin{bmatrix} \gamma w & \gamma \w^H \\ \gamma \w & \Z \end{bmatrix} = \U_r^H \R_{rr}^{-1} \U_r,
\end{equation}
we propose to use as initial point the first column of $\U_r^H \S_{rr}^{-1} \U_r,$
%\begin{equation}
%	\label{eq:initialization}
%	\U_r^H \S_{rr}^{-1} \U_r,
%\end{equation}
normalized to have unit norm.

\subsection{Generalized likelihood ratio test}

The generalized log-likelihood ratio is
\begin{align}
	\frac{1}{N} \log \Lambda &= \frac{1}{N} \log \ell(\hat{\Thetab},\hat{\gamma},\hat{\x},\hat{\Z}, \hat{q}_{sr}; \Y ) - \frac{1}{N} \log \ell (\hat{\R}_0 ; \Y) \nonumber \\
	&= \log \det(\S_{rr})- \log (\u_s^H \S_{ss}^{-1} \u_s)   \nonumber \\ &\phantom{=}  - \log \det \left(\V_r^H \S_{rr} \V_r\right) + \log \left(|\nu|^2\right),
		\label{eq:GLR_almost_final}
\end{align}
where $\hat{\x}$ is the maximizer of \eqref{eq:cost_function} obtained through numerical optimization and
\begin{equation}
	\label{eq:nu_squared}
	|\nu|^2 = \frac{\hat{\x}^H \E \hat{\x}}{\hat{\x}^H \Xib \hat{\x}} \frac{\hat{\x}^H \Psib \hat{\x}}{\hat{\x}^H \Gammab \hat{\x}}.
\end{equation}
The matrix $\Xib$ is defined in \eqref{eq:Xi_definition} and the matrices $\Psib$ and $\Gammab$  are defined in \eqref{eq:Psi_definition} and \eqref{eq:Gamma_definition}, respectively, of Appendix \ref{App:re_parametrization}.

Appendix \ref{App:simplification} derives a simpler expression for the GLR in \eqref{eq:GLR_almost_final}, which is given by
\begin{equation}
	\Lambda^{1/N} = \frac{|\nu|^2}{(\u_s^H\S_{ss}^{-1} \u_s ) (\u_r^H\S_{rr}^{-1} \u_r)}. 
	\label{eq:GLR_final}
\end{equation}

The algorithm, which is summarized in Alg. \ref{alg:alg_GLR}, has a computational complexity that is given by the computation of $\Psib, \Gammab,$ and $\Xib$, the maximization of $J(\x)$, and the evaluation of \eqref{eq:GLR_final}. The computation of $\Psib, \Gammab,$ and $\Xib$ reduces to products, inverses and sums of $L \times L$ matrices, which should be acceptable as $L$ is typically not very large. Moreover, the evaluation of \eqref{eq:GLR_final} is dominated by the computation of quadratic forms of $L \times L$ matrices. Then, the complexity of the algorithm is driven by the maximization of $J(\x)$, which depends on the chosen algorithm. In any case, it should be reasonable as the optimization variable is an $L$-dimensional vector.

\begin{algorithm}[!t]
	\caption{Computation of the GLR}
	\label{alg:alg_GLR}
	\begin{algorithmic}[1]
		\State {\bf Input}: $\S, \u_s \text{ and } \u_r$
		\State Obtain $\Psib, \Gammab,$ and $\Xib$ according to \eqref{eq:Psi_definition}, \eqref{eq:Gamma_definition}, and \eqref{eq:Xi_definition}
		\State Maximize $J(\x)$ in \eqref{eq:cost_function} using the proposed initialization
		\State Compute $\u_s^H\S_{ss}^{-1} \u_s$ and $\u_r^H\S_{rr}^{-1} \u_r$
		\State {\bf Output}:  $\Lambda^{1/N}$, which is defined in \eqref{eq:GLR_final}.
	\end{algorithmic}
\end{algorithm}

\section{Approximate GLRs}
\label{sec:simplfied}

This section presents two approximate detectors that do not require any numerical optimization. However, as the numerical simulations will show, their performance is competitive to that of  \eqref{eq:GLR_final} in many scenarios.

\subsection{Sample-based GLRT}

As we have seen in the previous section, maximizing \eqref{eq:log_likeli_Theta_b_qsr} with respect to $\R_{rr}$ is a complicated process with no closed-form solution. To avoid the numerical maximization, we propose to use a sample-based estimator for $\R_{rr}$ (instead of its ML estimator), which is given by $\hat{\R}_{rr}= \S_{rr}$. Using this sample-based estimate, the compressed log-likelihood is
\begin{multline}
	% \log \ell(\hat{\Thetab}_a, \hat{\Thetab},\hat{q}_{sr}; \Y ) =  - 2 L N (\log \pi + 1) - N \log \det(\S_{ss}) \\ - N \log \det(\S_{rr}) \\ + N \log \left(1 + \frac{|\eta_{sr}|^2}{\eta_s (\eta_r - \alpha)}\right),
	\log \ell(\hat{\Thetab},\hat{\R}_{rr},\hat{q}_{sr}; \Y ) =  - \log \det(\S_{ss}) - \log \det(\S_{rr}) \\ 
+ \log \left(1 + \frac{|\eta_{sr}(\S_{rr})|^2}{\u_s^H \S_{ss}^{-1} \u_s (\u_r^H \S_{rr}^{-1} \u_r - \alpha_{sr}(\S_{rr}))}\right).
\end{multline}

Finally, the sample-based likelihood ratio is
\begin{align}
  \Lambda_{app} &= \frac{\ell(\hat{\Thetab},\hat{\R}_{rr},\hat{q}_{sr}; \Y )}{\ell (\hat{\R}_0 ; \Y)} \nonumber \\
  &=   \left(1 + \frac{|\eta_{sr}(\S_{rr})|^2}{\u_s^H \S_{ss}^{-1} \u_s (\u_r^H \S_{rr}^{-1} \u_r - \alpha_{sr}(\S_{rr}))}\right)^N.
\end{align}
The proposed monotone function of $\Lambda_{app}$ is
\begin{equation}
 \label{eq:ICASSP_detector}
 \lambda_{app} = \Lambda_{app}^{1/N} - 1 = \frac{|\eta_{sr}(\S_{rr})|^2}{\u_s^H \S_{ss}^{-1} \u_s (\u_r^H \S_{rr}^{-1} \u_r - \alpha_{sr}(\S_{rr}))}. 
\end{equation}
More will be said about this detector in Section \ref{sec:interpretation}.

\subsection{Low-SNR GLRT}

For a low-SNR scenario, we can assume that \eqref{eq:R_1} is almost block-diagonal and hence estimate the diagonal blocks using the sample covariance matrices for the surveillance and reference channels. That is, \eqref{eq:R_1} could be estimated in a first step as
	\begin{equation}
	\label{eq:R_1_sample}
	\hat{\R}_1 =  \begin{bmatrix}  \S_{ss} & q_{sr} \u_s \u_r^H \\  q_{sr}^* \u_r \u_s^H  & \S_{rr} \end{bmatrix},
\end{equation}
where $q_{sr}$ is a parameter still to be estimated. The ML estimation of $q_{sr}$ can be formulated as \cite[Theorem 4.1]{Coherence_book}
\begin{equation}
	\label{eq:optimization_q_sr}
	\begin{array}{ll}
		\displaystyle \mathop{\text{minimize}}_{q_{sr} \in \mathbb{C}} \ &\det (\R_1) ,
		\\
		\mathop{\text{subject to}} \ &\tr \left(\R_1^{-1} \S \right) = 2L. 
	\end{array}
\end{equation}

To solve this problem, we need the determinant and inverse of \eqref{eq:R_1_sample}. The determinant is given by
	\begin{align}
	\det (\R_1) &= \det(\S_{ss}) \det \left(\S_{rr} - |q_{sr}|^2 \u_r \u_s^H \S_{ss}^{-1} \u_s \u_r^H \right) \nonumber \\
	&= \det(\S_{ss}) \det(\S_{rr}) \times \nonumber \\
	&\phantom{=} \left(1 - (\u_s^H\S_{ss}^{-1} \u_s ) (\u_r^H\S_{rr}^{-1} \u_r)|q_{sr}|^2\right), \label{eq:det_R_1_sample}
\end{align}
where we have used Schur's determinant identity and the matrix determinant lemma. This allows us to simplify \eqref{eq:optimization_q_sr} to
\begin{equation}
	\label{eq:optimization_q_sr_simp}
	\begin{array}{ll}
		\displaystyle \mathop{\text{maximize}}_{q_{sr} \in \mathbb{C}} \ & |q_{sr}|,\\
		\mathop{\text{subject to}} \ &\tr \left(\R_1^{-1} \S \right) = 2L.
	\end{array}
\end{equation}

Now, the trace can be written as
\begin{multline}
	\tr \left(\R_1^{-1} \S \right) = 2 L \ + \\ 2 \frac{(\u_s^H\S_{ss}^{-1} \u_s ) (\u_r^H\S_{rr}^{-1} \u_r) |q_{sr}|^2 - \Re\{q_{sr} \eta_{sr}^{\ast}(\S_{rr})\}}{1 - (\u_s^H\S_{ss}^{-1} \u_s ) (\u_r^H\S_{rr}^{-1} \u_r) |q_{sr}|^2},
\end{multline}
where we have used the expression for the inverse of a block matrix. The trace constraint therefore becomes
\begin{equation}
	(\u_s^H\S_{ss}^{-1} \u_s ) (\u_r^H\S_{rr}^{-1} \u_r) |q_{sr}|^2 - \Re\{q_{sr} \eta_{sr}^{\ast}(\S_{rr})\} = 0,
\end{equation}
allowing us to parametrize $q_{sr}$ as
\begin{equation}
	\label{eq:q_sr_parametric}
	q_{sr} = \frac{\eta_{sr}(\S_{rr}) + e^{j \theta} |\eta_{sr}(\S_{rr})|}{2 (\u_s^H\S_{ss}^{-1} \u_s ) (\u_r^H\S_{rr}^{-1} \u_r)} , \quad 0 \leq \theta < 2 \pi. 
\end{equation}
Using \eqref{eq:q_sr_parametric}, the solution to the optimization problem \eqref{eq:optimization_q_sr_simp} is 
\begin{equation}
	\hat{q}_{sr} = \frac{\eta_{sr}(\S_{rr}) }{(\u_s^H\S_{ss}^{-1} \u_s ) (\u_r^H\S_{rr}^{-1} \u_r)}, 
\end{equation}
and the compressed log-likelihood is
\begin{multline}
	% \log \ell (\hat{\R}_1 ; \Y) = - 2 L N (\log \pi + 1) - N \log \det(\S_{ss}) - N \log \det(\S_{rr}) \\ - N \log \left(1 - \frac{|\eta_{sr}|^2}{\eta_{s} \eta_{r}}\right),
	\log \ell (\hat{\R}_1 ; \Y) = - \log \det(\S_{ss}) - \log \det(\S_{rr}) \\ - \log \left(1 - \frac{|\eta_{sr}(\S_{rr})|^2}{(\u_s^H\S_{ss}^{-1} \u_s ) (\u_r^H\S_{rr}^{-1} \u_r)}\right).
\end{multline}
Finally, the likelihood ratio reduces to
\begin{equation}
	\Lambda_{low} = \frac{\ell (\hat{\R}_1 ; \Y)}{\ell (\hat{\R}_0 ; \Y)} 
	=   \left(1 + \frac{|\eta_{sr}(\S_{rr})|^2}{(\u_s^H\S_{ss}^{-1} \u_s ) (\u_r^H\S_{rr}^{-1} \u_r)}\right)^N,
\end{equation}
and the proposed low-SNR GLR is
\begin{equation}
	\label{eq:coherence}
	\lambda_{low} = \Lambda_{low}^{1/N} -  1= \frac{|\eta_{sr}(\S_{rr})|^2}{(\u_s^H\S_{ss}^{-1} \u_s ) (\u_r^H\S_{rr}^{-1} \u_r)}.
\end{equation}

\section{Interpretation and comparison of the detectors}
\label{sec:interpretation}

The detector in \eqref{eq:coherence} has an insightful interpretation. Defining the Capon beamformers for the surveillance and reference channels
\begin{align}
	\b_s &= \frac{\S_{ss}^{-1} \u_s}{\u_s^H \S_{ss}^{-1} \u_s}, & 	\b_r &= \frac{\S_{rr}^{-1} \u_r}{\u_r^H \S_{rr}^{-1} \u_r},
\end{align}
the statistic in \eqref{eq:coherence} can be rewritten as
\begin{equation}
	\lambda_{low} = \frac{|\u_s^H \S_{ss}^{-1} \S_{sr} \S_{rr}^{-1}\u_r|^2}{(\u_s^H \S_{ss}^{-1} \u_s)(\u_r^H \S_{rr}^{-1} \u_r)} = \frac{|\b_s^H \S_{sr} \b_r|^2}{(\b_s^H \S_{ss} \b_s)(\b_r^H \S_{rr} \b_r)}.
\end{equation}
So $\lambda_{low}$ is the magnitude-squared correlation coefficient between the outputs of the surveillance and reference Capon beamformers, which is a measure of coherence \cite{Coherence_book}.

Interestingly, $\lambda_{low}$ can also be expressed in terms of the normalized adaptive whitened matched filters. Defining these beamformers as
\begin{align}
	\w_s &= \frac{\S_{ss}^{-1/2} \u_s}{(\u_s^H \S_{ss}^{-1} \u_s)^{1/2}}, & \w_r &=  \frac{\S_{rr}^{-1/2} \u_r}{(\u_r^H \S_{rr}^{-1} \u_r)^{1/2}},
\end{align}
the low-SNR GLR is
\begin{equation}
	\lambda_{low} = | \w_s^H \C\w_r|^2,
\end{equation}
with the coherence matrix $\C = \S_{ss}^{-1/2} \S_{sr} \S_{rr}^{-1/2}$. Then, $\lambda_{low}$ can be interpreted as an adaptive, whitened matched filtering of coherence.

The detector in \eqref{eq:ICASSP_detector} also has an insightful interpretation based on the previous one. A few lines of algebra show that
\begin{equation}
	\lambda_{app} = \frac{\lambda_{low}}{1 - (\b_r^H \S_{rr} \b_r) \alpha_{sr}(\S_{rr})} = \frac{| \w_s^H \C\w_r|^2}{1 - \w_r^H \C^H \C \w_r}.
	\end{equation}
This statistic is quite intuitive. Each term is a normalized, whitened matched filtering of a coherence matrix or the coherence matrix squared. The denominator is bounded below by $0$ and above by $1$, so that $\lambda_{app}$ might be said to be an adaptive inflation of the statistic $\lambda_{low} = |\w_s^H\C \w_r|^2$.  So, in both cases, likelihood reasoning has produced a function of coherence, but this function is quite unlike the maximum singular value of coherence, $\sigma_{max}(\C)$, for detecting a subspace signal when only the dimension is known, as in \cite{Nacho_passive_rank_1}. It is not simply cross correlation. It is as if measurements $\Y_r$ in the reference channel are whitened as $\S_{rr}^{-1/2}\Y_r $ and measurements $\Y_s$ in the surveillance channel are whitened as $\S_{ss}^{-1/2}\Y_s$. These whitened measurements are cross correlated to form the coherence matrix $\C$. Only then is beamforming used, and it is beamforming by the whitened and normalized beamformers $\w_s$ and $\w_r$. It is not front-end beamforming with the conventional beamformers $\u_s$ and $\u_r$.

Finally, note the similarity between the new detectors of this paper, given by \eqref{eq:GLR_final}, \eqref{eq:ICASSP_detector} and \eqref{eq:coherence}, where the only difference is the terms $|\nu|^2$ (which appears instead of $|\eta_{sr}(\S_{rr})|^2$) and $\alpha_{sr}(\S_{rr})$ (which appears in the denominator of \eqref{eq:ICASSP_detector}). Then, we can say that the term $|\nu|^2$ plays the role of $|\eta_{sr}(\S_{rr})|^2$, when $\R_{rr}$ is estimated using the ML estimator. That is, the numerator of \eqref{eq:GLR_final} measures, in a different manner, the cross-covariance in the known signal subspaces between surveillance and reference channels, and the denominator acts as a normalization term that considers the individual channels. Another common property to all three detectors is that they are constant false alarm rate (CFAR) detectors to independent scalings of the two channels.

\begin{figure}[!t]
	\centering
	\includegraphics[width=0.95\columnwidth]{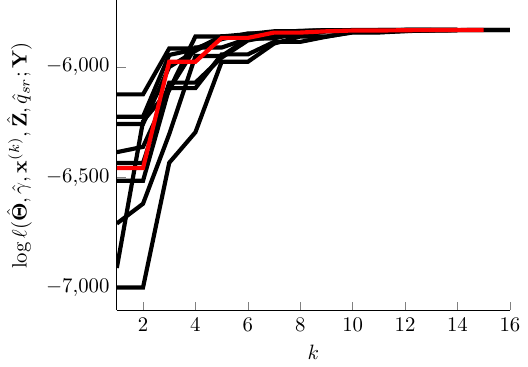}
	\caption{Convergence of the log-likelihood under $\mathcal{H}_1$ for an experiment with $N = 200, L = 8, \text{SNR}_r = 10$ dBs, and $\text{SNR}_s = 0$ dBs}
	\label{fig:convergence}
\end{figure}

\section{Numerical results}
\label{sec:simulations}

In this section, we study, by means of Monte Carlo simulations, the performances of the proposed detectors and compare them to the performance of the GLR for unknown channels \cite{Nacho_passive_rank_1}. This GLR  is the largest canonical correlation coefficient $\sigma_{max}(\C)$, which is the largest singular value of the coherence matrix $\C = \S_{ss}^{-1/2} \S_{sr} \S_{rr}^{-1/2}$. Moreover, we have also included in the comparison the cross-correlation detector \cite{Himed_cross_correlation}, $T_{cc} = \|\S_{sr}\|^2_{F}$, and the detector \cite{gogineni_passive_2018} $T_{svd} = |\v_s^H \v_r|^2$, where $\v_s$ and $\v_r$ are, respectively, the dominant right\footnote{We must point out that in \cite{gogineni_passive_2018,Nadakuditi_passive}, the left singular vectors are used since they consider the ``transpose'' model.} singular vectors of $\Y_s$ and $\Y_r$. Additionally, we also analyze the convergence of the numerical approach used to optimize \eqref{eq:cost_function} and the null distributions of the GLR. 

\subsection{Convergence}

For a fixed set of observations generated under $\mathcal{H}_1$, we study the convergence of MATLAB's \verb|fminunc|'s trust-region algorithm \cite{MATLAB_fminunc} to optimize \eqref{eq:cost_function}. In particular, we consider a scenario with $N = 200, L = 8,$ and $\text{SNR}_r = 10$ dBs, $\text{SNR}_s = 0$ dBs, with the SNR defined as the input signal-to-noise ratio
\begin{equation}
	\text{SNR}_i = 10 \log_{10}\left( \frac{\sigma_x^2 \|\h_i\|^2}{\tr(\Sigmab_{ii})}\right),
\end{equation}
where $i = \{s,r\}$. Fig. \ref{fig:convergence} shows the value of the compressed log-likelihood $\log \ell(\hat{\boldsymbol{\Theta}},\hat{\gamma},\mathbf{x}^{(k)},\hat{\mathbf{Z}},\hat{q}_{sr}; \mathbf{Y} )$, where $\x^{(k)}$ is the solution at iteration $k$, for $9$ random initializations (shown in black) and for the initialization based on the first column of $\U_r^H \S_{rr}^{-1} \U_r$. This experiment is just a representative case of the usual behavior, where the proposed initialization achieves a (possibly local) maximum in a few iterations. Random initializations typically achieve the same value of the log-likelihood as the proposed initialization, but may require more iterations.

In the following experiments, different channels and noise covariance matrices are generated for each Monte Carlo simulation, and we have considered $10^6$ trials.  The channel gains are proper complex Gaussians, which yield amplitudes that are Rayleigh distributed, and the noise covariance matrices are Wishart distributed, which are scaled to achieve a desired SNR. 

\subsection{ROC curves}

\begin{figure}[!t]
	\centering
	\includegraphics[width=0.95\columnwidth]{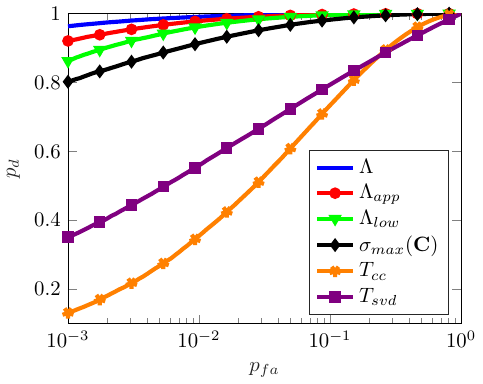}
	\caption{ROC curves for an experiment with $N = 15, L = 4, \text{SNR}_r = 15$ dBs, and $\text{SNR}_s = -5$ dBs}
	\label{fig:ROC}
\end{figure}

The second experiment compares the receiver operating characteristic (ROC) curves of the proposed detectors in \eqref{eq:GLR_final}, \eqref{eq:ICASSP_detector}, and \eqref{eq:coherence}, the GLR for unknown channels \cite{Nacho_passive_rank_1}, $\sigma_{max}(\C)$, the cross-correlation detector, $T_{cc}$, and the SVD detector, $T_{svd}$. In particular, Fig. \ref{fig:ROC} depicts the ROC curves for an experiment with $N = 15, L = 4, \text{SNR}_r = 15$ dBs, and $\text{SNR}_s = -5$ dBs, where we can see that \eqref{eq:GLR_final} presents the best performance, followed closely by \eqref{eq:ICASSP_detector} and \eqref{eq:coherence}. Additionally, this figure also shows the performance improvement due to knowing the channel subspaces. The performances of $T_{cc}$ and $T_{svd}$ are relatively poor for this scenario, as neither exploits known array response vectors and neither accounts for general positive definite noise covariances.

\subsection{Probability of missed detection}

\begin{figure}[!t]
	\centering
	\includegraphics[width=0.95\columnwidth]{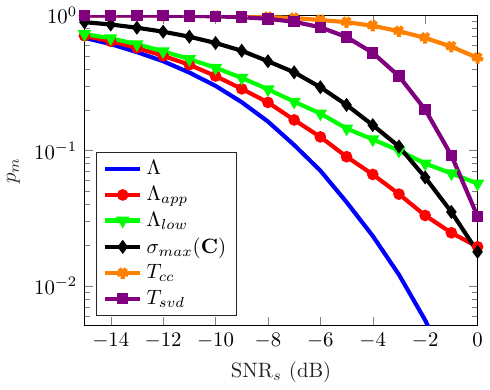}
	\caption{Probability of missed detection vs. $\text{SNR}_s$ for an experiment with $N = 15, L = 4, \text{SNR}_r = \text{SNR}_s + 10$ dBs, and $p_{fa} = 10^{-3}$}
	\label{fig:pm_SNR_10}
\end{figure}

The next experiment evaluates the probability of missed detection ($p_m$) for $p_{fa} = 10^{-3}$ and varying $\text{SNR}_s$ in a scenario with $N = 15, \text{SNR}_r = \text{SNR}_s + 10$ dBs, and $L = 4$. Fig. \ref{fig:pm_SNR_10} shows the results for the three proposed detectors, where we can see that for low SNRs all detectors perform similarly. However, as the SNR grows, the GLR in \eqref{eq:GLR_final} performs better, followed closely by the approximate GLR in \eqref{eq:ICASSP_detector}, at least for moderate SNRs. However, for larger SNRs, there is a clear advantage for the GLR in \eqref{eq:GLR_final}. Additionally, the three remaining detectors present, in general, a poor performance, as they do not use the channel subspace. Nonetheless, for large SNR, the performance of \cite{Nacho_passive_rank_1} surpasses that of \eqref{eq:coherence}, which was derived for low SNRs, and so does $T_{svd}$.

\begin{figure}[!t]
	\centering
	\includegraphics[width=0.95\columnwidth]{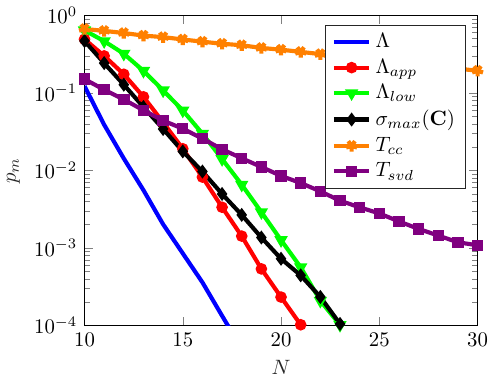}
	\caption{Probability of missed detection vs. $N$ for an experiment with $L = 4, \text{SNR}_r = 10$ dBs, $\text{SNR}_s = 0$ dBs, and $p_{fa} = 10^{-3}$}
	\label{fig:pm_N}
\end{figure}

\begin{figure}[!t]
	\centering
	\includegraphics[width=0.95\columnwidth]{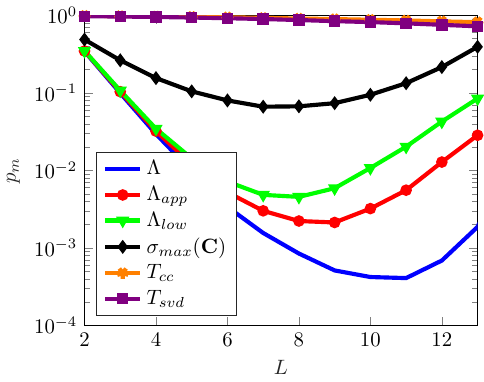}
	\caption{Probability of missed detection vs. $L$ for an experiment with $N = 30, \text{SNR}_r = 0$ dBs, $\text{SNR}_s = -10$ dBs, and $p_{fa} = 10^{-3}$}
	\label{fig:pm_L}
\end{figure}

Fig. \ref{fig:pm_N} depicts the probability of missed detection vs. $N$ for an experiment with $L = 4, \text{SNR}_r = \text{SNR}_s = 0$ dBs, and $p_{fa} = 10^{-3}$, where again \eqref{eq:GLR_final} presents the best performance, over a wide range of values of $N$, followed by \eqref{eq:ICASSP_detector} and \eqref{eq:coherence}. Again, the detectors that do not use the channel subspace present bad performance,  as expected since they only exploit the dimension of the reference and surveillance subspaces, and not their spans. Additionally, Fig. \ref{fig:pm_L} shows the probability of missed detection vs. $L$ for an experiment with $N = 30, \text{SNR}_r = \text{SNR}_s = -10$ dBs, and $p_{fa} = 10^{-3}$, where similar conclusions can be drawn. Interestingly, for the proposed detectors, there is a value of $L$ for which the performance worsens (if $N$ is fixed). Finally, we must point out that for low SNR and large number of snapshots (e.g., $N=500$), other simulations indicate that the GLR and its approximations have nearly identical performances.

\subsection{Null distribution}

\begin{figure}[!t]
	\centering
	\includegraphics[width=0.95\columnwidth]{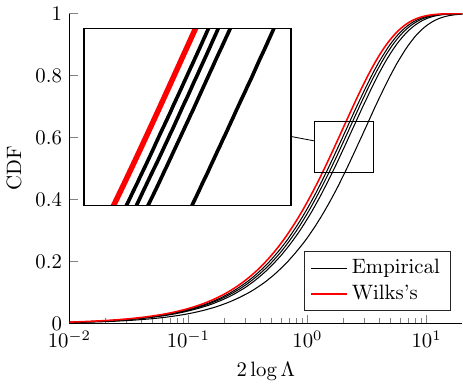}
	\caption{Cumulative distribution function (CDF) under $\mathcal{H}_0$ for an experiment with $L = 4, \text{SNR}_r = 0$ dBs. The red curve is the CDF of Wilks's approximation and the black curves correspond to the empirical CDF obtained for $N = 20, 40, 60, $ and $100$ (the curves for larger $N$ are closer to the CDF of Wilks's approximation)}
	\label{fig:distributions}
\end{figure}

Finally, we study the distribution of the detector in \eqref{eq:GLR_final} based on Wilks's theorem \cite{Box_wilks}. For large $N$, the distribution of the GLR is approximated as the chi-squared distribution
\begin{equation}
	2 \log \Lambda \mathop{\sim}^{N \rightarrow \infty} \chi^2_2.
\end{equation}
This is a chi-squared distribution with two degrees of freedom, which is an exponential distribution, since the difference in the number of free parameters between both hypotheses is two (due to the complex $q_{sr}$). Fig. \ref{fig:distributions} shows the results for an experiment with $L = 4$ and $\text{SNR}_r = 0$ dBs, and depicts the cumulative distribution function (CDF) obtained through Wilks's theorem and estimated using Monte Carlo simulations. It is clear that, even for small values of $N$, the approximation given by a $\chi^2_2$ distribution is accurate in the body of the distribution.

\section{Conclusions}
\label{sec:conclusions}

A source of opportunity transmits signals to a multisensor reference array that receives them in additive Gaussian noise of unknown covariance. The question is whether this transmission is also received in a multisensor surveillance array in additive noise of unknown covariance. It is assumed that each sensor array is modeled by its known array response vector, but path loss, shadowing, fading, and carrier phase to each array are unknown. So the received signal at each multisensor array consists of a sequence of visits to a known subspace. Several detectors are derived from a generalized likelihood principle and its approximations. Of these, one is an ``exact'' GLRT based on a numerical optimization and the other two are computationally attractive functions of squared coherence between Capon beamformed measurements. Each of these outperforms detectors based on cross-correlations between measurements at the reference and surveillance arrays. A comparison between the detectors of this paper and the singular-value detector of \cite{Nacho_passive_rank_1} for the case where the array response vectors are unknown quantifies the value of knowing them.

\appendices

\renewcommand{\thesection}{\Alph{section}.\arabic{section}}
\setcounter{section}{0}

\begin{appendices}
	
\section{Proof of Lemma \ref{lem:ML_Theta_b_q_sr}}
\label{App:ML_Theta_b_q_sr}

The log-likelihood in \eqref{eq:loglikeli_Stoica} is composed of two terms: 1) a Gaussian log-likelihood with covariance matrix $\Thetab$ and sample covariance matrix $\M(q_{sr},\R_{rr})$; and 2) another Gaussian log-likelihood with covariance matrix $\R_{rr}$ and sample covariance matrix $\S_{rr}$. It is tempting to estimate $\R_{rr}$ and $\Thetab$ independently, but it is not possible because $\M(q_{sr},\R_{rr})$ depends on $\R_{rr}$. However, the first two terms in \eqref{eq:loglikeli_Stoica} do not depend on $\Thetab$, which allows us to write \cite{mardia_multivariate_analysis}
\begin{align}
	\hat{\Thetab} &= \arg \min_{\Thetab} \ \log \det(\Thetab) + \tr \left(\Thetab^{-1} \M(q_{sr},\R_{rr}) \right) \nonumber \\ &= \M(q_{sr},\R_{rr}).
\end{align}
Then, the compressed log-likelihood in $\Thetab$ is
\begin{multline}
	\label{eq:log_likeli_Theta_b}
	% \max_{\Thetab} \log \ell (\R_{rr}, \Thetab, q_{sr} ; \Y) = \log \ell(\R_{rr}, \hat{\Thetab},q_{sr}; \Y ) =  - L N (2\log \pi + 1) - N \log \det(\R_{rr}) \\ - N \tr \left(\R_{rr}^{-1} \S_{rr} \right) - N \log \det(\M(q_{sr},\R_{rr})).
	\max_{\Thetab} \, \log \ell (\Thetab,\R_{rr}, q_{sr} ; \Y) =\log \ell(\hat{\Thetab},\R_{rr}, q_{sr}; \Y ) =  \\ - \log \det(\R_{rr}) - \tr \left(\R_{rr}^{-1} \S_{rr} \right) - \log \det(\M(q_{sr},\R_{rr})).
\end{multline}
The next step in the proof addresses the maximization in $q_{sr}$. The compressed log-likelihood \eqref{eq:log_likeli_Theta_b} depends on $q_{sr}$ only through $\M(q_{sr},\R_{rr})$, and hence
\begin{equation}
	\hat{q}_{sr} = \arg \min_{q_{sr}} \quad \, \det(\M(q_{sr},\R_{rr})).
\end{equation}
The matrix $\M(q_{sr},\R_{rr})$ in \eqref{eq:M_matrix} may be factored as
\begin{multline}
	\M(q_{sr},\R_{rr}) = \S_{ss}^{1/2} \left[ \I_{L} + |q_{sr}|^2 \eta_r(\R_{rr}) \S_{ss}^{-1/2} \u_s  \u_s^H \S_{ss}^{-1/2} \right. \\-  q_{sr} \S_{ss}^{-1/2} \u_s \u_r^H \R_{rr}^{-1} \S_{sr}^H \S_{ss}^{-1/2} \\  \left.   - q_{sr}^{\ast} \S_{ss}^{-1/2} \S_{sr} \R_{rr}^{-1} \u_r \u_s^H \S_{ss}^{-1/2} \right] \S_{ss}^{1/2},
\end{multline}
which can be rewritten as
\begin{equation}
	\M(q_{sr},\R_{rr}) = \S_{ss}^{1/2} \left( \I_{L} +  \F \G \F^H  \right) \S_{ss}^{1/2},
\end{equation}
where
\begin{equation}
	\F = \begin{bmatrix}
		\S_{ss}^{-1/2} \u_s  & \S_{ss}^{-1/2} \S_{sr} \R_{rr}^{-1} \u_r
	\end{bmatrix} \in \mathbb{C}^{L \times 2},
\end{equation}
and
\begin{equation}
	\G = \begin{bmatrix}
		|q_{sr}|^2 \eta_r(\R_{rr}) & - q_{sr} \\ - q_{sr}^{\ast} & 0
	\end{bmatrix}.
\end{equation}
Hence, the determinant may be written
\begin{align}
	\det(\M(q_{sr},\R_{rr})) &=  \det(\S_{ss}) \det\left( \I_{L} +  \F \G \F^H \right) \nonumber \\ &= \det(\S_{ss}) \det\left( \I_{2} +  \F^H \F \G  \right),
\end{align}
where we have used the matrix determinant lemma. A few straightforward operations show that 
\begin{multline}
	\det\left( \I_{2} +  \F^H \F \G  \right) =  1 - \eta_{sr}^{\ast}(\R_{rr}) q_{sr} - \eta_{sr}(\R_{rr}) q_{sr}^{\ast}  \\  + |q_{sr}|^2 \left[(\u_s^H \S_{ss}^{-1} \u_s)  \eta_r(\R_{rr}) + |\eta_{sr}(\R_{rr})|^2 \right. \\ \left. - (\u_s^H \S_{ss}^{-1} \u_s)  \alpha_{sr}(\R_{rr}) \right].
\end{multline}
Complete the square to get \eqref{eq:q_sr_hat}.

\section{Derivation of \eqref{eq:log_likeli_Theta_b_qsr_reparametrized}}
\label{App:re_parametrization}

This appendix rewrites \eqref{eq:log_likeli_Theta_b_qsr} in terms of the new parameters, $\gamma,\x,$ and $\Z$, starting with
\begin{multline}
	-\log \det(\R_{rr}) = \log \det(\X) = \log w + \log \gamma \\ + \log \det \left(\Z -  \frac{\gamma}{w} \w \w^H\right),
\end{multline}
where we have used Schur's determinant identity. The trace term is
\begin{multline}
	\tr \left(\R_{rr}^{-1} \S_{rr} \right) = \gamma w \u_r^H\S_{rr} \u_r + 2 \gamma \Re \left\{ \w^H \V_r^H\S_{rr} \u_r \right\} \\ + \tr\left(\Z \V_r^H \S_{rr} \V_r\right).
\end{multline}
The last term of \eqref{eq:log_likeli_Theta_b_qsr} is
\begin{multline}
	\log \left(\u_s^H \S_{ss}^{-1} \u_s + \frac{|\eta_{sr}(\R_{rr})|^2}{\eta_r(\R_{rr}) - \alpha(\R_{rr})}\right) = \\ \log \left( \u_s^H \S_{ss}^{-1} \u_s + \frac{\u_r^H \R_{rr}^{-1} \left(\S_{sr}^H \S_{ss}^{-1} \u_s  \u_s^H \S_{ss}^{-1} \S_{sr} \right) \R_{rr}^{-1} \u_r}{\u_r^H\R_{rr}^{-1} \left( \S_{rr} - \S_{sr}^H \S_{ss}^{-1} \S_{sr} \right) \R_{rr}^{-1} \u_r}\right) \\ = \log \left(\frac{ \x^H \Psib \x}{\x^H \Gammab \x}\right),
\end{multline}
where 
\begin{multline}
	\label{eq:Psi_definition}
	\Psib = \U_r^H \left[(\u_s^H \S_{ss}^{-1} \u_s)  \left( \S_{rr} - \S_{sr}^H \S_{ss}^{-1} \S_{sr} \right) \right. \\ \left. + \, \S_{sr}^H \S_{ss}^{-1} \u_s  \u_s^H \S_{ss}^{-1} \S_{sr} \right] \U_r,
\end{multline}
and
\begin{equation}
	\label{eq:Gamma_definition}
	\Gammab = \U_r^H \left( \S_{rr} - \S_{sr}^H \S_{ss}^{-1} \S_{sr} \right) \U_r.
\end{equation}
We have used the identity
\begin{equation}
	\R_{rr}^{-1} \u_r = \U_r \X \U_r^H \u_r = \gamma \U_r \x.
\end{equation}
Taking the above expressions into account, \eqref{eq:log_likeli_Theta_b_qsr} becomes \eqref{eq:log_likeli_Theta_b_qsr_reparametrized}.

\section{Proof of Lemma \ref{lem:estimates_Z_gamma}}
\label{App:estimates_Z_gamma}

The maximization in $\Z$ of the log-likelihood \eqref{eq:log_likeli_Theta_b_qsr_reparametrized} is equivalent to
\begin{equation}
	\label{eq:log_likeli_Theta_b_qsr_reparametrized_noconstant}
	\mathop{\text{maximize}}_{\Z} \, \log \det \left(\Z - \frac{\gamma}{w} \w \w^H\right)  - \tr\left(\Z \V_r^H \S_{rr} \V_r\right) .
\end{equation}
To solve the previous optimization problem, let us define $\B^{-1} = \Z - \frac{\gamma}{w} \w \w^H$ and write \eqref{eq:log_likeli_Theta_b_qsr_reparametrized_noconstant} as
\begin{align}
	&\log \det \left(\B^{-1}\right) - \tr\left[\left(\B^{-1} +  \frac{\gamma}{w} \w \w^H\right) \V_r^H \S_{rr} \V_r\right] \nonumber \\
	=  &-\log \det \left(\B\right) - \tr\left(\B^{-1} \V_r^H \S_{rr} \V_r \right) - \frac{\gamma}{w} \w^H \V_r^H \S_{rr} \V_r \w.
\end{align}
The first two terms, which are the only ones that depend on $\B$, are proportional to a Gaussian log-likelihood with covariance matrix $\B$ and sample covariance matrix $\V_r^H \S_{rr} \V_r$. Therefore, assuming that $\B$ is only constrained as $\B \succ \0$, the maximizer is $\hat{\B} = \V_r^H \S_{rr} \V_r$, which yields the ML estimate for $\Z$ in \eqref{eq:ML_Z}. However, $\B$ is not unconstrained since $\Z \succ \0$ and $w - \gamma \w^H \Z^{-1} \w > 0$. Then, we must check whether $\hat{\Z}$ satisfies these constraints. First, it is clear that $\hat{\Z} \succ \0$, so it suffices to check the second constraint. To do so, let us use the Sherman-Morrison identity to obtain the inverse of $\hat{\Z}$ as 
\begin{equation}
	\hat{\Z}^{-1} = \V_r^H \S_{rr} \V_r - \frac{\V_r^H \S_{rr} \V_r \w  \w^H \V_r^H \S_{rr} \V_r}{\frac{w}{\gamma} + \w^H \V_r^H \S_{rr} \V_r \w}, 
\end{equation}
which yields
\begin{equation}
	\w^H \hat{\Z}^{-1} \w = \frac{w \w^H \V_r^H \S_{rr} \V_r \w}{w + \gamma \w^H \V_r^H \S_{rr} \V_r \w}.
\end{equation}
Then, the condition is fulfilled:
\begin{equation}
	w - \gamma  \w^H \hat{\Z}^{-1} \w = \frac{w^2}{w + \gamma \w^H \V_r^H \S_{rr} \V_r \w} > 0,
\end{equation}
since $w, \gamma >0$ and $\w^H \V_r^H \S_{rr} \V_r \w \geq 0$. Plugging back $\hat{\Z}$ into \eqref{eq:log_likeli_Theta_b_qsr_reparametrized}, the compressed log-likelihood becomes
\begin{multline}
	\label{eq:log_likeli_Theta_b_qsr_Z}
	%\log \ell(\gamma,\x,\hat{\Z}, \hat{\Thetab},\hat{q}_{sr}; \Y ) =  - 2 L N ( \log \pi + 1) + N - N \log \det(\S_{ss}) \\ - N \log (\eta_s) + N\log \gamma + N \log w - N\log \det \left(\V_r^H \S_{rr} \V_r\right) \\  - N \gamma w \u_r^H\S_{rr} \u_r - 2 N \gamma \Re \left\{ \w^H \V_r^H\S_{rr} \u_r \right\} \\ - N \frac{\gamma}{w} \w^H\V_r^H \S_{rr} \V_r \w  + N \log \left(\frac{ \x^H \Psib \x}{\x^H \Gammab \x}\right),
	\log \ell(\hat{\Thetab},\gamma,\x,\hat{\Z}, \hat{q}_{sr}; \Y ) =  - \log \det(\S_{ss})  - \log (\u_s^H \S_{ss}^{-1} \u_s)  \\ + \log \gamma +  \log w - \log \det \left(\V_r^H \S_{rr} \V_r\right) - \gamma w \u_r^H\S_{rr} \u_r \\ - 2 \gamma \Re \left\{ \w^H \V_r^H\S_{rr} \u_r \right\} - \frac{\gamma}{w} \w^H\V_r^H \S_{rr} \V_r \w  + \log \left(\frac{ \x^H \Psib \x}{\x^H \Gammab \x}\right).
\end{multline}
The derivative of \eqref{eq:log_likeli_Theta_b_qsr_Z} is
\begin{multline}
	\label{eq:partial_gamma}
	\frac{\partial \log \ell(\hat{\Thetab},\gamma,\x,\hat{\Z}, \hat{q}_{sr}; \Y )}{\partial \gamma } = \frac{1}{\gamma}  -  w \u_r^H\S_{rr} \u_r \\ - 2 \Re \left\{ \w^H \V_r^H\S_{rr} \u_r \right\} - \frac{1}{w} \w^H\V_r^H \S_{rr} \V_r \w,
\end{multline}
and setting \eqref{eq:partial_gamma} to zero yields \eqref{eq:ML_gamma}, where we have taken into account that
\begin{multline}
	\x^H \Xib \x = w^2 \u_r^H\S_{rr} \u_r \\ + 2 w \Re \left\{ \w^H \V_r^H\S_{rr} \u_r \right\} +  \w^H \V_r^H \S_{rr} \V_r \w.
\end{multline}

\section{Derivation of \eqref{eq:GLR_final}}
\label{App:simplification}

This appendix simplifies \eqref{eq:GLR_almost_final} to get the more insightful detection statistic of \eqref{eq:GLR_final}. It is clear that
\begin{multline}
	\det(\S_{rr}) = \det(\Xib) = \det \left(\V_r^H \S_{rr} \V_r\right) \times \\ \left[\u_r^H\S_{rr} \u_r - \u_r^H\S_{rr} \V_r (\V_r^H \S_{rr} \V_r)^{-1} \V_r^H\S_{rr} \u_r\right],
\end{multline}
and therefore
\begin{multline}
	\beta = \frac{\det(\S_{rr})}{\det \left(\V_r^H \S_{rr} \V_r\right)} = \u_r^H\S_{rr} \u_r \\ - \u_r^H\S_{rr} \V_r (\V_r^H \S_{rr} \V_r)^{-1} \V_r^H\S_{rr} \u_r .
\end{multline}
But $\beta$ is the inverse of the NW block of $\Xib^{-1}$:
\begin{align}
	\Xib^{-1} &= \begin{bmatrix}   \u_r^H\S_{rr} \u_r & \u_r^H\S_{rr} \V_r  \\  \V_r^H\S_{rr} \u_r & \V_r^H \S_{rr} \V_r  \end{bmatrix}^{-1} \nonumber \\
	&=  \begin{bmatrix}   \beta^{-1} & NE  \\  SW & SE  \end{bmatrix}.
\end{align}
By the unitarity of $\U_r$, this inverse can also be computed as
\begin{equation}
	\Xib^{-1} = \U_r^H \S_{rr}^{-1} \U_r =  \begin{bmatrix}   \u_r^H\S_{rr}^{-1} \u_r & NE  \\  SW & SE  \end{bmatrix},
\end{equation}
yielding $\u_r^H\S_{rr}^{-1} \u_r =  \beta^{-1}$, from which it follows that
\begin{equation}
	\frac{\det(\S_{rr})}{\det \left(\V_r^H \S_{rr} \V_r\right)} = \frac{1}{\u_r^H\S_{rr}^{-1} \u_r}, 
\end{equation}
and the GLR becomes \eqref{eq:GLR_final}. 

\end{appendices}

\bibliographystyle{IEEEtran}
\bibliography{biblio}

\begin{thebibliography}{10}
\providecommand{\url}[1]{#1}
\csname url@rmstyle\endcsname
\providecommand{\newblock}{\relax}
\providecommand{\bibinfo}[2]{#2}
\providecommand\BIBentrySTDinterwordspacing{\spaceskip=0pt\relax}
\providecommand\BIBentryALTinterwordstretchfactor{4}
\providecommand\BIBentryALTinterwordspacing{\spaceskip=\fontdimen2\font plus
\BIBentryALTinterwordstretchfactor\fontdimen3\font minus
  \fontdimen4\font\relax}
\providecommand\BIBforeignlanguage[2]{{%
\expandafter\ifx\csname l@#1\endcsname\relax
\typeout{** WARNING: IEEEtran.bst: No hyphenation pattern has been}%
\typeout{** loaded for the language `#1'. Using the pattern for}%
\typeout{** the default language instead.}%
\else
\language=\csname l@#1\endcsname
\fi
#2}}

\bibitem{Griffiths_passive_radar_1}
H.~D. Griffiths and C.~J. Baker, ``Passive coherent location radar systems.
  {P}art 1: Performance prediction,'' \emph{IEE Proc. Radar, Sonar and
  Navigation}, vol. 152, no.~3, p. 153, 2005.

\bibitem{Griffiths_passive_radar_2}
C.~J. Baker, H.~D. Griffiths, and I.~Papoutsis, ``Passive coherent location
  radar systems. {P}art 2: Waveform properties,'' \emph{IEE Proc. Radar, Sonar
  and Navigation}, vol. 152, no.~3, p. 160, 2005.

\bibitem{Blum_receive_passive}
Y.~Li, Q.~He, and R.~S. Blum, ``Limited-complexity receiver design for
  passive/active {MIMO} radar detection,'' \emph{IEEE Trans. Signal Process.},
  vol.~67, no.~12, pp. 3258--3271, 2019.

\bibitem{passive_preamble}
Y.~Liu, G.~Liao, J.~Xu, Z.~Yang, and Y.~Yin, ``Improving detection performance
  of passive {MIMO} radar by exploiting the preamble information of
  communications signal,'' \emph{IEEE Systems Journal}, vol.~15, no.~3, pp.
  4391--4402, 2021.

\bibitem{Hack_passive_noreference_signal}
D.~E. Hack, L.~K. Patton, B.~Himed, and M.~A. Saville, ``Centralized passive
  {MIMO} radar detection without direct-path reference signals,'' \emph{IEEE
  Trans. Signal Process.}, vol.~62, no.~11, pp. 3013--3023, 2014.

\bibitem{Nuria_passive_radar}
A.~Ali, N.~Gonz{\'a}lez-Prelcic, and A.~Ghosh, ``Passive radar at the roadside
  unit to configure millimeter wave vehicle-to-infrastructure links,''
  \emph{IEEE Trans. Vehicular Tech.}, vol.~69, no.~12, pp. 14\,903--14\,917,
  2020.

\bibitem{DOA_passive_radar}
J.~Shen, X.~Wan, J.~Yi, W.~Zhang, and S.~Hu, ``{DOA} estimation for passive
  radar under the scene where target echo is partially correlated,'' \emph{IEEE
  Trans. Vehicular Tech.}, vol.~72, no.~4, pp. 4863--4874, 2023.

\bibitem{passive_radar_automative}
G.~P. Blasone, F.~Colone, and P.~Lombardo, ``Forward-looking passive radar with
  non-uniform linear array for automotive applications,'' \emph{IEEE Trans.
  Vehicular Tech.}, 2023.

\bibitem{Willis2007}
N.~J. Willis and H.~D. Griffiths, \emph{Advances in Bistatic Radar}.\hskip 1em
  plus 0.5em minus 0.4em\relax SciTech Publishing, 2007.

\bibitem{DVB_passive}
M.~Baczyk and M.~Malanowski, ``Reconstruction of the reference signal in
  {DVB-T}-based passive radar,'' \emph{Int. J. of Electronics and
  Telecommunications}, vol.~57, no.~1, pp. 43--48, 2011.

\bibitem{FM_passive}
P.~Howland, D.~Maksimiuk, and G.~Reitsma, ``{FM} radio based bistatic radar,''
  \emph{IEE Proc. Radar, Sonar and Navigation}, vol. 152, no.~3, pp. 107--115,
  2005.

\bibitem{Hack_passive_radar_MIMO_networks}
D.~E. Hack, L.~K. Patton, B.~Himed, and M.~A. Saville, ``Detection in passive
  {MIMO} radar networks,'' \emph{IEEE Trans. Signal Process.}, vol.~62, no.~11,
  pp. 2999--3012, 2014.

\bibitem{Scharf_PSL_first}
L.~T. McWhorter, L.~Scharf, C.~Moore, and M.~Cheney, ``Passive multi-channel
  detection: A general first-order statistical theory,'' \emph{IEEE Open J.
  Signal Process.}, vol.~4, pp. 437--451, 2023.

\bibitem{Zaimbashi_unified_2021}
A.~Zaimbashi, ``A unified framework for multistatic passive radar target
  detection under uncalibrated receivers,'' \emph{IEEE Trans. Signal Process.},
  vol.~69, pp. 695--708, 2021.

\bibitem{Coherence_book}
D.~Ram{\'\i}rez, I.~Santamar{\'\i}a, and L.~Scharf, \emph{Coherence: In Signal
  Processing and Machine Learning}.\hskip 1em plus 0.5em minus 0.4em\relax
  Springer Nature, 2023.

\bibitem{Colone_passive_2009}
F.~Colone, D.~W. O'Hagan, P.~Lombardo, and C.~J. Baker, ``A multistage
  processing algorithm for disturbance removal and target detection in passive
  bistatic radar,'' \emph{IEEE Trans. Aero. Electr. Syst.}, vol.~45, no.~2, pp.
  698--722, 2009.

\bibitem{Himed_cross_correlation}
J.~Liu, H.~Li, and B.~Himed, ``On the performance of the cross-correlation
  detector for passive radar applications,'' \emph{Signal Process.}, vol. 113,
  pp. 32--37, 2015.

\bibitem{Cui2015}
G.~Cui, J.~Liu, H.~Li, and B.~Himed, ``Signal detection with noisy reference
  for passive sensing,'' \emph{Signal Process.}, vol. 108, pp. 389--399, 2015.

\bibitem{Bialkowski2011}
K.~S. Bialkowski, I.~V.~L. Clarkson, and S.~D. Howard, ``Generalized canonical
  correlation for passive multistatic radar detection,'' in \emph{IEEE Work.
  Stat. Signal Process.}, {J}ul. 2011, pp. 417--420.

\bibitem{gogineni_passive_2018}
S.~Gogineni, P.~Setlur, M.~Rangaswamy, and R.~R. Nadakuditi, ``Passive radar
  detection with noisy reference channel using principal subspace similarity,''
  \emph{IEEE Trans. Aero. Electr. Syst.}, vol.~54, no.~1, pp. 18--36, 2018.

\bibitem{Nadakuditi_passive}
------, ``Random matrix theory inspired passive bistatic radar detection with
  noisy reference signal,'' in \emph{IEEE Int. Conf. on Acoustics, Speech and
  Signal Process.}, 2015.

\bibitem{Nacho_passive_rank_1}
I.~Santamar{\'{i}}a, L.~L. Scharf, D.~Cochran, and J.~V{\'{i}}a, ``Passive
  detection of rank-one signals with a multiantenna reference channel,'' in
  \emph{European Signal Process. Conf.}, {A}ug. 2016, pp. 140--144.

\bibitem{Nacho_passive_rank_p}
I.~Santamar{\'{i}}a, L.~Scharf, J.~Via, Y.~Wang, and H.~Wang, ``Passive
  detection of correlated subspace signals in two {MIMO} channels,'' \emph{IEEE
  Trans. Signal Process.}, vol.~65, no.~7, pp. 1752--1764, {M}ar. 2017.

\bibitem{Wang_Asilomar}
Y.~Wang, L.~L. Scharf, I.~Santamaria, and H.~Wang, ``Canonical correlations for
  target detection in a passive radar network,'' in \emph{Asilomar Conf.
  Signals, Systems, and Computers}, Pacific Grove, USA, {N}ov. 2016.

\bibitem{euspico17_passive}
I.~Santamaria, J.~V{\'{i}}a, L.~L. Scharf, and Y.~Wang, ``A {GLRT} approach for
  detecting correlated signals in white noise in two {MIMO} channels,'' in
  \emph{European Signal Process. Conf.}, {A}ug. 2017.

\bibitem{cochran_Trans_SP_95}
D.~Cochran, H.~Gish, and D.~Sinno, ``A geometric approach to multiple-channel
  signal detection,'' \emph{IEEE Trans. Signal Process.}, vol.~43, no.~9, pp.
  2049--2057, {S}ep. 1995.

\bibitem{yu_novel_2023}
T.~Song and J.~Zhu, ``A novel weighted-distance centralized detection method in
  passive {MIMO} radar,'' in \emph{Green, {Pervasive}, and {Cloud}
  {Computing}}.\hskip 1em plus 0.5em minus 0.4em\relax Springer International
  Publishing, 2023, vol. 13744, pp. 192--203, series Title: Lecture Notes in
  Computer Science.

\bibitem{javidan_target_2020}
M.~J. Hassan, A.~Zaimbashi, and J.~Liu, ``Target detection in passive radar
  under noisy reference channel: {A} new threshold-setting strategy,''
  \emph{IEEE Trans. Aero. Electr. Syst.}, vol.~56, pp. 4711--4722, {D}ec. 2020.

\bibitem{fazlollahpoor_rao_2020}
M.~Fazlollahpoor, M.~Derakhtian, and S.~Khorshidi, ``Rao retector for passive
  {MIMO} radar with direct-path interference,'' \emph{IEEE Trans. Aero. Electr.
  Syst.}, vol.~56, no.~4, pp. 2999--3009, {A}ug. 2020.

\bibitem{zhang_passive_2019}
X.~Zhang, F.~Wang, H.~Li, J.~Sward, A.~Jakobsson, and B.~Himed, ``Passive
  multistatic detection by exploiting a sparsity structure of the {IO}
  waveform,'' in \emph{IEEE Radar Conf.}, Boston, MA, USA, {A}pr. 2019.

\bibitem{zhang_sparsity-based_2019}
X.~Zhang, J.~Sward, H.~Li, A.~Jakobsson, and B.~Himed, ``A sparsity-based
  passive multistatic detector,'' \emph{IEEE Trans. Aero. Electr. Syst.},
  vol.~55, no.~6, pp. 3658--3666, {D}ec. 2019.

\bibitem{zhang_multistatic_2017}
X.~Zhang, H.~Li, and B.~Himed, ``Multistatic detection for passive radar with
  direct-path interference,'' \emph{IEEE Trans. Aero. Electr. Syst.}, vol.~53,
  no.~2, pp. 915--925, {A}pr. 2017.

\bibitem{liu_passive_2020}
Y.~Liu, R.~S. Blum, G.~Liao, and S.~Zhu, ``Passive {MIMO} radar detection
  exploiting known format of the communication signal observed in colored noise
  with unknown covariance matrix,'' \emph{Signal Process.}, vol. 174, p.
  107611, {S}ep. 2020.

\bibitem{Himed_MAP_radar_2021}
S.~Guruacharya, B.~K. Chalise, and B.~Himed, ``{MAP} ratio test detector for
  radar system,'' \emph{IEEE Trans. Signal Process.}, vol.~69, pp. 573--588,
  2021.

\bibitem{zhang_multistatic_2017-1}
X.~Zhang, H.~Li, and B.~Himed, ``Multistatic passive detection with parametric
  modeling of the {IO} waveform,'' \emph{Signal Processing}, vol. 141, pp.
  187--198, {D}ec. 2017.

\bibitem{chalise_glrt_2018}
B.~K. Chalise and B.~Himed, ``{GLRT} detector in single frequency multi-static
  passive radar systems,'' \emph{Signal Process.}, vol. 142, pp. 504--512,
  {J}an. 2018.

\bibitem{zaimbashi_multistatic_2020}
A.~Zaimbashi, ``Multistatic passive radar sensing algorithms with calibrated
  receivers,'' \emph{IEEE Sensors Journal}, vol.~20, no.~14, pp. 7878--7885,
  {J}ul. 2020.

\bibitem{liu_passive_2021}
Y.~Liu, G.~Liao, H.~Li, S.~Zhu, Y.~Li, and Y.~Yin, ``Passive {MIMO} radar
  detection with unknown colored {Gaussian} noise,'' \emph{Remote Sensing},
  vol.~13, no.~14, p. 2708, {J}ul. 2021.

\bibitem{fazlollahpoor_passive_2021}
M.~Fazlollahpoor, M.~Derakhtian, and S.~Khorshidi, ``Passive {MIMO} radar
  detection in the presence of clutter or multi‐path without reference
  channel,'' \emph{IEE Proc. Radar, Sonar and Navigation}, vol.~15, no.~2, pp.
  154--166, {F}eb. 2021.

\bibitem{zaimbashi_multistatic_2022}
A.~Zaimbashi and M.~S. Greco, ``Multistatic passive radar target detection
  under uncalibrated receivers with direct-path interference,'' \emph{IEEE
  Trans. Aero. Electr. Syst.}, vol.~58, no.~6, pp. 5443--5455, {D}ec. 2022.

\bibitem{pak_target_2019}
S.~Pak, B.~K. Chalise, and B.~Himed, ``Target localization in multi-static
  passive radar systems with artificial neural networks,'' in \emph{IEEE Radar
  Conf.}, {S}ep. 2019, pp. 1--5.

\bibitem{wang_signal_2018}
F.~Wang, H.~Li, X.~Zhang, and B.~Himed, ``Signal parameter estimation for
  passive bistatic radar with waveform correlation exploitation,'' \emph{IEEE
  Trans. Aero. Electr. Syst.}, vol.~54, no.~3, pp. 1135--1150, {J}un. 2018.

\bibitem{wang_joint_2019}
L.~Wang, Q.~He, R.~S. Blum, and H.~Li, ``Joint parameter estimation employing
  coherent passive {MIMO} radar,'' \emph{J. of Engineering}, vol. 2019, no.~20,
  pp. 6859--6862, {O}ct. 2019.

\bibitem{chen_direct_2023}
Y.~Chen, P.~Wei, H.~Zhang, M.~You, and W.~Li, ``Direct target joint detection
  and tracking based on passive multi-static radar,'' \emph{Remote Sensing},
  vol.~15, no.~3, p. 624, {J}an. 2023.

\bibitem{chalise_performance_2017}
B.~K. Chalise, M.~G. Amin, and B.~Himed, ``Performance tradeoff in a unified
  passive radar and communications system,'' \emph{IEEE Signal Process. Lett.},
  vol.~24, no.~9, pp. 1275--1279, {S}ep. 2017.

\bibitem{Rank_one_known_ICASSP}
D.~Ram{\'\i}rez, I.~Santamar{\'\i}a, and L.~Scharf, ``Passive detection of
  rank-one {G}aussian signals for known channel subspaces and arbitrary
  noise,'' in \emph{IEEE Int. Conf. on Acoustics, Speech and Signal Process.},
  2023.

\bibitem{MATLAB_fminunc}
T.~F. Coleman and Y.~Li, ``An interior trust region approach for nonlinear
  minimization subject to bounds,'' \emph{SIAM J. Opt.}, vol.~6, no.~2, pp.
  418---445, {M}ay 1996.

\bibitem{Zhao_linear_fusion_2017}
H.-Y. Zhao, J.~Liu, Z.-J. Zhang, H.~Liu, and S.~Zhou, ``Linear fusion for
  target detection in passive multistatic radar,'' \emph{Signal Process.}, vol.
  130, pp. 175--182, 2017.

\bibitem{Zhang_delay_Doppler_direct_2016}
X.~Zhang, H.~Li, J.~Liu, and B.~Himed, ``Joint delay and {D}oppler estimation
  for passive sensing with direct-path interference,'' \emph{IEEE Trans. Signal
  Process.}, vol.~64, no.~3, pp. 630--640, 2016.

\bibitem{Ramirez_MFA_2020}
D.~Ram{\'i}rez, I.~Santamar{\'i}a, L.~L. Scharf, and S.~V. Vaerenbergh,
  ``Multi-channel factor analysis with common and unique factors,'' \emph{IEEE
  Trans. Signal Process.}, vol.~68, pp. 113--126, 2020.

\bibitem{first_vs_second_TSP}
I.~Santamar{\'\i}a, L.~L. Scharf, and D.~Ram{\'\i}rez, ``Scale-invariant
  subspace detectors based on first- and second-order statistical models,''
  \emph{IEEE Trans. Signal Process.}, vol.~68, pp. 6432--6443, 2020.

\bibitem{Kay_detection}
S.~M. Kay, \emph{Fundamentals of Statistical Signal Processing: Detection
  Theory}.\hskip 1em plus 0.5em minus 0.4em\relax Prentice Hall, 1998, vol.~II.

\bibitem{book_peter}
P.~J. Schreier and L.~L. Scharf, \emph{Statistical Signal Processing of
  Complex-Valued Data}.\hskip 1em plus 0.5em minus 0.4em\relax Cambridge
  University Press, 2010.

\bibitem{mardia_multivariate_analysis}
K.~V. Mardia, J.~T. Kent, and J.~M. Bibby, \emph{Multivariate Analysis}.\hskip
  1em plus 0.5em minus 0.4em\relax New York: Academic, 1979.

\bibitem{Stoica_array_correlared}
P.~Stoica, K.-M. Wong, and Q.~Wu, ``On a nonparametric detection method for
  array signal processing in correlated noise fields,'' \emph{IEEE Trans.
  Signal Process.}, vol.~44, no.~4, pp. 1030--1032, {A}pr. 1996.

\bibitem{Box_wilks}
G.~E.~P. Box, ``A general distribution theory for a class of likelihood
  criteria,'' \emph{Biometrika}, vol.~36, pp. 317--346, {D}ec. 1949.

\end{thebibliography}

\end{document}